\documentclass[runningheads,11pt,envcountsame]{llncs}
\usepackage[T1]{fontenc}
\usepackage[margin=1in]{geometry}

\usepackage{graphicx}
\usepackage{microtype}
\usepackage{amsmath}
\usepackage{amssymb}
\usepackage{dsfont}
\usepackage{comment}

\usepackage{amsthm}
\usepackage{mathtools}
\usepackage[createShortEnv]{proof-at-the-end}

\newif\iflong
\longtrue 

\iflong 
\pgfkeys{/prAtEnd/proofEnd/.style={normal}}
\pgfkeys{/prAtEnd/allEnd/.style={normal}}
\pgfkeys{/prAtEnd/textEnd/.style={both}}
\else
\pgfkeys{/prAtEnd/proofEnd/.style={end, restate, no link to proof, text proof}}
\pgfkeys{/prAtEnd/allEnd/.style={all end,  text proof}}
\pgfkeys{/prAtEnd/textEnd/.style={end}}
\fi

\usepackage{multirow}
\usepackage{ctable}
\usepackage
[
    subrefformat = simple, 
    labelformat = simple,  
]
{subcaption}

\usepackage[hidelinks]{hyperref}
\usepackage{cleveref}

\spnewtheorem{ourconjecture}{Conjecture}{\bfseries}{\itshape}

\newcommand*{\R}{\mathds{R}}
\newcommand*{\s}{\mathbf{s}}
\DeclareMathOperator*{\argmax}{arg\,max}
\DeclareMathOperator*{\argmin}{arg\,min}

\usepackage{color}

\begin{document}
%

\title{Cooperation in Bilateral Generalized Network Creation}

%
%
\author{\href{https://orcid.org/0000-0001-8394-4469}{Hans Gawendowicz}\inst{1} \and
\href{https://orcid.org/0000-0002-3010-1019}{Pascal Lenzner}\inst{2} \and
Lukas Weyand\inst{3}}

%
\institute{Hasso-Plattner-Institut Potsdam, Germany, 
\email{hans.gawendowicz@hpi.de}\and
University of Augsburg, Germany,
\email{pascal.lenzner@uni-a.de}\and
Hasso-Plattner-Institut Potsdam, Germany, 
\email{lukas.weyand@student.hpi.de}}

\maketitle              
\begin{abstract}
Studying the impact of cooperation in strategic settings is one of the cornerstones of algorithmic game theory. Intuitively, allowing more cooperation yields equilibria that are more beneficial for the society of agents. However, for many games it is still an open question how much cooperation is actually needed to ensure socially good equilibria. 
We contribute to this research endeavor by analyzing the benefits of cooperation in a network formation game that models the creation of communication networks via the interaction of selfish agents. In our game, agents that correspond to nodes of a network can buy incident edges of a given weighted host graph to increase their centrality in the formed network. The cost of an edge is proportional to its length, and both endpoints must agree and pay for an edge to be created. This setting is known for having a high price of anarchy. 

To uncover the impact of cooperation, we investigate the price of anarchy of our network formation game with respect to multiple solution concepts that allow for varying amounts of cooperation. On the negative side, we show that on host graphs with arbitrary edge weights even the strongest form of cooperation cannot improve the price of anarchy. In contrast to this, as our main result, we show that cooperation has a significant positive impact if the given host graph has metric edge weights. For this, we prove asymptotically tight bounds on the price of anarchy via a novel proof technique that might be of independent interest and can be applied in other models with metric weights.

\end{abstract}

\addtocounter{page}{-1}
\section{Introduction}
With the rise of the Internet, the study of networks has become a prime research topic that proved to be essential for understanding today's world~\cite{barabasi2013network}. The study of network optimization has a long tradition in computer science and operations research, but classical combinatorial optimization relies on having a central authority that can shape and govern the network. This might be feasible on a small scale, but our most important networks, like the Internet, 
have long outgrown this scale. Modern large-scale communication networks are shaped by the interaction of many decentralized entities, e.g. Internet service providers, which usually are independent for-profit firms that act selfishly in order to maximize their own benefit. This calls for an analysis via algorithmic game theory~\cite{Pap01}. 

Since roughly three decades, various game-theoretic models for network creation have been studied. In these games, agents corresponding to nodes in a network strategically form edges among themselves. Any strategy profile then corresponds to a created network, and the objects of study are the created networks that correspond to equilibria of the underlying game. The archetypical example of such a model is the \emph{network creation game (NCG)} by Fabrikant, Luthra, Maneva, Papadimitriou, and Shenker~\cite{NCG_OG}, which was recently extended to the \emph{generalized network creation game (GNCG)}~\cite{GNCG}. The latter better models communication networks, e.g. fiber-optic networks, where the distance between nodes plays a key role by allowing the creation of edges with different lengths/latencies. 

Typically, the focus of the analysis of game-theoretic network creation is on how the selfish action of the agents affects the overall quality of the created networks. This is measured via the price of anarchy (PoA)~\cite{PoA}, which compares the overall cost of the worst possible equilibrium network created by the interaction of selfish agents with the cost of the best network obtained by centralized optimization.  
However, even after decades of intensive research, many intriguing questions about the PoA of game-theoretic network creation models are still open. One open problem is the trade-off between the amount of allowed cooperation among the agents and the obtained PoA. Intuitively, the more the agents cooperate, the lower should be the impact of selfishness on the created networks. But how much cooperation is actually needed to achieve a significant improvement of the PoA? Besides being theoretically interesting, answering this question also has a significant impact as it serves as guideline for policy-makers and regulating authorities.

This question has been studied for the original NCG and variants~\cite{different_cooperative_network_creation_games,NCG_equilibrium_non-existence}, where edges can be unilaterally created by any endpoint. Moreover, very recently it was studied for the \emph{bilateral network creation game (BNCG)}~\cite{CooperationBNCG}, where consent of both endpoints is needed for creating an edge. 

In this paper, we set out to tackle the question of how cooperation influences the PoA for the much more intricate \emph{bilateral generalized network creation game}. While it is known that the PoA in both the GNCG and the BNCG can be very high, as our main result, we show that both a high amount of cooperation and a geometric setting, where the edge lengths in the network are metric, are necessary  for significantly improving the PoA.  

\subsection{Model and Notation}
\label{section:model_notation}
We consider the \emph{bilateral generalized network creation game} (BGNCG), a bilateral version of the GNCG~\cite{GNCG}. Let $\mathcal{G}$ be the set of all BGNCG instances. An instance $(H, \alpha) \in \mathcal{G}$ is given as a complete and undirected weighted host graph $H = (V, E_H, w)$ and a fixed value of the parameter $\alpha \in \R^+$, where $V$ is a set of $n$ nodes, $E_H$ is the set of edges in $H$. The parameter $\alpha$ allows for modeling a trade-off between the cost of buying edges and the ability to use them. The edges in $H$ have arbitrary non-negative edge weights $w\colon E_H \rightarrow \R_0^+$ and for $u,v\in V$, we will use the shortcut $w(u,v)$ for $w(\{u,v\})$. Additionally, we write $w(M)$ to denote $\sum_{\{u,v\} \in M} w(u,v)$, where $M \subseteq E_H$. 

A \emph{strategy} $S_u$ of agent $u$ is a subset of $V \setminus \{u\}$ that determines which other agents agent~$u$ wants to build an edge to. If $v \in S_u$, for some $v \in V$, then agent~$u$ has to pay the cost of the edge $\{u,v\}$. This cost is proportional to weight $w(u,v)$ of the edge, in particular, we set it to $\alpha \cdot w(u,v)$. Furthermore, let $\s = (S_{v_1},...,S_{v_n})$ be the \emph{strategy profile}, i.e., the vector containing the strategies of all agents. The strategy profile $\s$ defines a subgraph of $H$ we denote as $G(\s) = (V, E(\s))$, where $E(\s) = \{ \{u,v\} \mid v \in S_u \wedge u \in S_v \}$. Note that since the game is bilateral, an edge $\{u,v\}$ is only part of $G(\s)$ if both agent~$u$ and agent~$v$ want to build it. For this reason, we will only consider strategy profiles where either $v \in S_u \wedge u \in S_v$ or $v \notin S_u \wedge u \notin S_v$, as all other strategy profiles are unstable since some agent pays for an edge that is not built. Hence, we have a bijection between strategy profiles and undirected graphs, and we will use these representations interchangeably. 

Let $d_G(u,v)$ be the \emph{distance between $u$ and $v$ in $G$}, i.e., the length of the shortest path between $u$ and $v$ in $G$ with respect to the sum of the edge weights of all edges on that path, or $\infty$ if no such path exists. For convenience, we define $d_G(u, V) \coloneqq \sum_{v \in V} d_G(u,v)$ and $w(u,S_u) \coloneqq \sum_{v \in S_u} w(u,v)$. We refer to $d_G(u, V)$ as the \emph{distance cost of $u$}, and to $\alpha \cdot w(u,S_u)$ as the \emph{edge cost of $u$}. Given a strategy profile $\s$, the total cost of agent $u$ in $G(\s)$ is then defined as
$$cost(u, G(\s)) = \alpha \cdot w(u,S_u) + d_{G(\s)}(u, V).$$

All agents act selfishly and strategically try to minimize their own cost. Whether agent $u$ deviates from her current strategy $S_u$ to another strategy $S'_u$ depends on the \emph{solution concept} used. We consider a variety of distinct solution concepts (defined below) that specify whether a graph $G(\s)$ is considered \emph{stable}, i.e., whether there does not exist a set of agents that can improve their cost by changing their strategy within the limitations of the respective solution concept. If there indeed is such a set of agents that can improve in such a way, we call the corresponding joint strategy change an \emph{improving move}. In particular, we focus on solution concepts that allow for different degrees of cooperation between the agents. In the BGNCG, removing an edge can be done unilaterally, but when adding one or more edges, all the agents who have to modify their strategy accordingly to enable this have to benefit from the change. We consider the following known solution concepts:
\begin{itemize}
    \item[$\bullet$] \textbf{Pairwise Stability (PS)~\cite{jackson2003strategic}:} A strategy profile $\s$ is \emph{pairwise stable (PS)}, if
    \begin{enumerate}
    \item[(1)] no agent $u\in V$ can decrease her cost by removing a single node from $S_u$ and
    \item[(2)] if no two agents $u,v \in V$ can decrease their cost by adding $u$ to $S_v$ and $v$ to $S_u$.\footnote{In PS, every edge is wanted by both endpoints and every non-edge is not wanted by at least one of the endpoints.}
    \end{enumerate}
    \item[$\bullet$] \textbf{Bilateral Neighborhood Equilibrium (BNE)~\cite{CooperationBNCG}:} A strategy profile $\s$ is in \emph{bilateral neighborhood equilibrium (BNE)} if there is no agent $u \in V$ with the following type of improving move: Let $R \subseteq S_u$ and let $A \subseteq V \setminus S_u$. Removing every node of $R$ from $S_u$ and for every $v\in A$ adding $v$ to $S_u$ and $u$ to $S_v$. Agent $u$ and all agents in $A$ strictly decrease their individual cost.\footnote{The BNE is the bilateral analogue of the Nash equilibrium for unilateral games. The only difference is, that for all created edges, the respective endpoints have to jointly agree (and pay) for these edges.}
    \item[$\bullet$] \textbf{Bilateral Strong Equilibrium (BSE)\cite{CooperationBNCG}:} A strategy profile $\s$ is in \emph{bilateral strong equilibrium} (BSE) if there is no coalition $\Gamma \subseteq V$ such that there is the following type of improving move: For every $u \in \Gamma$ removing every node of a subset $R_u \subseteq S_u$ from $S_u$. At the same time, for every $u \in \Gamma$ adding every node of a subset $A_u \subseteq \Gamma \setminus \{u\}$ to $S_u$ and adding $u$ to $S_v$ for all $v \in A_u$. All agents in $\Gamma$ strictly decrease their individual cost.\footnote{Since edges can only be created bilaterally, the BSE is equivalent to the well-known strong Nash equilibrium~\cite{NCG_equilibrium_non-existence}.}
\end{itemize}
 Note, that by definition every strategy profile in BSE is also in BNE and every strategy profile in BNE is also pairwise stable.

 We measure the impact of selfishness via the \emph{price of anarchy} (PoA) of the above solution concepts. For this, we first define the \emph{social cost} of a network $G(\s)$, denoted $cost(G(\s))$, as $\sum_{u \in V} cost(u,G(\s))$, i.e., the total cost of all agents. The subgraph of $H$ that minimizes the social cost among all possible strategy profiles is called the \emph{social optimum network} $OPT_{H, \alpha}=(V, E_{OPT_{H, \alpha}})$. If $(H,\alpha)$ is clear from the context, we just denote the social optimum network as $OPT$.
 Let $\mathcal{G}_{\alpha,n}$ be the set of all instances with $n$ nodes and parameter $\alpha$ and let $worst_{H, \alpha}$ be the equilibrium network with the highest social cost for a given instance $(H,\alpha)$ and given solution concept $X$. Then the price of anarchy with respect to $X$ is defined as $$PoA(\alpha, n)\coloneqq\max_{(H,\alpha)\in \mathcal{G}_{n,\alpha}}  \frac{cost(worst_{H, \alpha})}{cost(OPT_{H, \alpha})}.$$ 
 We often write $PoA$ instead of $PoA(\alpha, n)$.

 From now on, we mostly abstract away from strategy profiles and work directly with their corresponding networks, i.e., writing $G=(V,E)$ instead of $G(\s)=(V,E(\s))$. In addition, we write $G-R$ and $G+R$, where $R \subseteq E_H$, to denote the graphs $(V, E \setminus R)$ and $(V, E \cup R)$, respectively.
 
We also consider a natural model variation of the BGNCG with metric edge weights. In the \emph{metric variant of the BGNCG} (\emph{M-BGNCG}) the edge weights must satisfy the triangle inequality, that is, for any nodes $u,v,z \in V$ it holds that $w(u,v) \leq w(u,z) + w(z,v)$.

\subsection{Related Work}
The NCG by Fabrikant, Luthra, Maneva, Papadimitriou, and Shenker~\cite{NCG_OG} is well-known in the field of game theoretic network formation models. A sequence of works over two decades~\cite{NCG_OG,NCG_BFS_tree,MihalakS13,MamageishviliMM15,BiloL20,AlvarezM19} has established that the PoA of the NCG is constant for almost the whole range of the edge price parameter $\alpha$, while the best general PoA upper bound of $2^{O(\sqrt{\log n})}$ was obtained by Demaine, Hajiaghayi, Mahini, and Zadimoghaddam~\cite{DemaineHMZ12}. The NCG has influenced many other game-theoretic network formation models. Since it is infeasible to list all of the research on these models, we focus on the models that are similar to our model and that share some key features. Therefore, we focus on models with edge-weighted host graphs or models that incorporate cooperation between the agents. Furthermore, we limit the discussion of these models to their PoA.

Our model is based on the GNCG by Bilò, Friedrich, Lenzner, and Melnichenko~\cite{GNCG}, which is the first variant of the NCG that incorporates weighted edges. They show that the PoA of the GNCG is at least $\frac{\alpha+2}{2}$ while Friedemann, Friedrich, Gawendowicz, Lenzner, Melnichenko, Peters, Stephan, and Vaichenker~\cite{Efficiency_and_Stability_in_Euclidean_Network_Design} contribute an upper bound of $2(\alpha+1)$. Thus, the PoA of the GNCG is in $\Theta(\alpha)$. This is in stark contrast to the PoA of the NCG. Another contrast is that the existence of Nash equilibria for the GNCG is an open problem, only known for restricted edge weights, e.g., weights given by a tree metric. For our model, equilibrium existence is also an open problem. 
Other than the GNCG, there also are other models with non-uniform edge costs \cite{edge_cost_graph_dependent,edge_cost_pop_based,edge_interval_buy,edge_cost_internet_based,Bilo0LLM21}, but they either do not consider edges of varying lengths or incorporate quite different types of possible strategies.

There are several network creation games that consider cooperation between agents. 
The bilateral version of the NCG, i.e., the BNCG, employs the concept of pairwise stability~\cite{jackson2003strategic} and was introduced by Corbo and Parkes~\cite{BNCG_OG_bounds}. For this, a tight bound of $\Theta(\sqrt{\alpha})$ for $\alpha \leq n$ and $\Theta(n/\sqrt{\alpha})$ for $\alpha > n$ on the PoA is known~\cite{BNCG_OG_bounds,DemaineHMZ12}. With closer inspection, the lower bound actually is $\Omega ( \frac{n\sqrt{\alpha}}{n+\alpha} )$. 

Further degrees of cooperation are considered by Friedrich, Gawendowicz, Lenzner, and Zahn~\cite{CooperationBNCG}, who also introduce the solution concepts BNE and BSE. Similar to our approach, they analyze the PoA of the BNCG for different amounts of cooperation. However, besides working with a complete unweighted host graph, their work differs significantly from our approach since they focus on the analysis of equilibrium networks which are trees. They show that just additionally allowing cooperative swaps of edges results in a PoA of $\Theta(\log(\alpha))$ for stable tree networks. Moreover, for tree networks in BNE, they show a PoA of $\Theta(\log(\alpha))$ if $\alpha \geq n^{1/2+ \epsilon}$, and a PoA of $\Theta(1)$ if $\alpha \leq \sqrt{n}$. Furthermore, they prove a PoA of $\Theta(1)$ for tree networks in $3$-BSE, which is similar to the BSE but only coalitions of size at most $3$ are allowed. For general networks in BSE, they show that the PoA is $\Theta(1)$ if $\alpha \leq n^{1- \epsilon}$ or if $\alpha \geq n \log(n)$, and that the PoA is in $\mathcal{O} ( \frac{\log(n)}{\log \log \log(n)})$ for all values of $\alpha$. 

Finally, also the original NCG~\cite{NCG_OG}, which is a special case of the GNCG where the given host graph is unweighted and complete, has been analyzed with respect to BSE as well. For the original NCG, the PoA for networks in strong Nash equilibrium~\cite{aumann1959acceptable}, the unilateral version of the BSE, for $\alpha \geq 2$ is at most $2$ \cite{NCG_equilibrium_non-existence}, and it is at most $\frac{3}{2}$~\cite{JanusK17}. Interestingly, it was shown that strong Nash equilibria do not exist for $1<\alpha < 2$. The latter is addressed by the study of near-strong equilibria, a weaker version of strong Nash equilibria, which do exist for some allowed coalition sizes~\cite{RozenfeldT10}. Moreover, also a cooperative network creation model~\cite{different_cooperative_network_creation_games}, where agents can buy cost-shares of any edge, was studied. Due to the possibility of buying non-incident edges, this model is fundamentally different.

\subsection{Our Contribution}
We introduce the bilateral generalized network creation game (BGNCG), a bilateral version of the generalized network creation game~\cite{GNCG}, where a given weighted host graph defines the lengths of all edges. For this intricate model and the natural special case with metric weights, we analyze the PoA with respect to different solution concepts that allow for varying amounts of cooperation among the agents. \Cref{fig:PoA_table} gives an overview of our results.

\begin{table}[ht]
\centering
\setlength{\tabcolsep}{5.9pt} 
\renewcommand{\arraystretch}{1.4} 
\begin{tabular}{ l l l l l }
    \specialrule{1.5pt}{0pt}{0pt}
    Model & $\alpha$ & PS & BNE & BSE \\
    \specialrule{1.5pt}{2pt}{0pt}
    \multirow{6}{*}{\textbf{M-BGNCG}} & \multirow{2}{*}{$\alpha \leq n$} & $\Omega(\alpha)$ [Thm.\,\ref{M-BGNCG_PoA_lower_bounds}] & $\Omega \left( \sqrt{\alpha} \right)$ [Thm.\,\ref{M-BGNCG_PoA_lower_bounds}] & $\Omega \left( \sqrt{\alpha} \right)$ [Thm.\,\ref{M-BGNCG_PoA_lower_bounds}] \\
    & & $\leq \alpha + 1$ [Thm.\,\ref{PoA_upper_M-BGNCG_PS_alpha}] & $\leq \alpha + 1$ [Thm.\,\ref{PoA_upper_M-BGNCG_PS_alpha}] & $\mathcal{O} \left( \sqrt{\alpha} \right)$ [Cor.\,\ref{PoA_upper_M-BGNCG_BSE}] \\
    \cline{2-5}
    & \multirow{2}{*}{$n \leq \alpha \leq n^2$} & $\Omega(n)$ [Thm.\,\ref{M-BGNCG_PoA_lower_bounds}] & $\Omega \left( \sqrt{\alpha} \right)$ [Thm.\,\ref{M-BGNCG_PoA_lower_bounds}] & $\Omega \left( \max \left\{ \frac{n}{\sqrt{\alpha}}, \frac{\alpha}{n} \right\} \right) $ [Thm.\,\ref{M-BGNCG_PoA_lower_bounds}] \\
    & & $\leq 2n$ [Thm.\,\ref{PoA_upper_M-BGNCG_PS_n}] & $\leq 2n$ [Thm.\,\ref{PoA_upper_M-BGNCG_PS_n}] & $\mathcal{O} \left( \min \left\{ \frac{\alpha\sqrt{\alpha}}{n}, 2n \right\} \right)$ [Thms.\,\ref{PoA_upper_M-BGNCG_BSE}, \ref{PoA_upper_M-BGNCG_PS_n}]\\
    \cline{2-5}
    & \multirow{2}{*}{$\alpha \geq n^2$} & $\Omega(n)$ [Thm.\,\ref{M-BGNCG_PoA_lower_bounds}] & $\Omega(n)$ [Thm.\,\ref{M-BGNCG_PoA_lower_bounds}] & $\Omega (n)$ [Thm.\,\ref{M-BGNCG_PoA_lower_bounds}] \\
    & & $\leq 2n$ [Thm.\,\ref{PoA_upper_M-BGNCG_PS_n}] & $\leq 2n$ [Thm.\,\ref{PoA_upper_M-BGNCG_PS_n}] & $\leq 2n$ [Thm.\,\ref{PoA_upper_M-BGNCG_PS_n}] \\
    \specialrule{1.5pt}{2pt}{0pt}
    \multirow{2}{*}{\textbf{BGNCG}} & \multirow{2}{*}{all} & $\geq \alpha + 1$ [Thm.\,\ref{PoA_lower_BGNCG_BSE_alpha}] & $\geq \alpha + 1$ [Thm.\,\ref{PoA_lower_BGNCG_BSE_alpha}] & $\geq \alpha + 1$ [Thm.\,\ref{PoA_lower_BGNCG_BSE_alpha}] \\
    & & $\leq 2(\alpha + 1)$ [Thm.\,\ref{PoA_upper_BGNCG_PS_alpha}] & $\leq 2(\alpha + 1)$ [Thm.\,\ref{PoA_upper_BGNCG_PS_alpha}] & $\leq 2(\alpha + 1)$ [Thm.\,\ref{PoA_upper_BGNCG_PS_alpha}] \\
    \specialrule{1.5pt}{2pt}{0pt}
\end{tabular}
\caption{Result overview. Lower and upper bounds one the PoA of the BGNCG and the M-BGNCG for multiple solution concepts that allow varying degrees of cooperation.}
\label{fig:PoA_table}
\end{table}

 For arbitrary edge weights of the given host graph, that is, for the unrestricted BGNCG, we show that the PoA is linear in $\alpha$ for all considered solution concepts. This mirrors the respective results for the GNCG. For the M-BGNCG, where the edge weights must satisfy the triangle inequality, the PoA does eventually improve if we increase the allowed amount of cooperation.
 
 Our main result is the upper bound of $\mathcal{O} ( \min \{ \frac{\alpha\sqrt{\alpha}}{n}, 2n \} )$ on the PoA of M-BGNCG networks in BSE, which, for $\alpha \in o ( n^{4/3})$, is a significant improvement over the upper bound on the PoA for M-BGNCG networks in BNE. Most notably, for $\alpha \in \mathcal{O}(n)$, this results in a tight $\Theta \left( \sqrt{\alpha} \right)$ upper bound on the PoA of M-BGNCG networks in BSE, since we provide a matching lower bound of $\Omega ( \frac{n\sqrt{\alpha}}{n+\alpha})$. We derive these results with entirely novel techniques that might be interesting in their own right, since they open up a new approach for PoA bounds in cooperative network formation models. Moreover, our improved bounds for the BSE versus the BNE show that for successful cooperation, besides having a geometric setting, not only the consent of other agents (agreement to proposed edges) is important but rather their active participation (joint proposal of new edges to be formed).  


\section{Preliminaries}\label{sec:prelim}
We introduce several technical statements that hold for the BGNCG and that provide important structural insights into equilibrium networks. 

The first insight is that an agent that would improve when multiple of its incident edges are removed from the network, can also improve by removing one of those edges. This also holds for the BNCG, as Corbo and Parkes \cite{BNCG_OG_bounds} show. We present a novel proof for our more general setting.

\begin{lemmaE}[][proofEnd, category=prelim]
\label{remove_mult_remove_one}
    Let $G=(V,E)$ be a BGNCG network. If removing multiple edges incident to some agent~$u \in V$ in $G$ results in a lower total cost for agent~$u$, then there exists an improving move for agent $u$ in $G$ that only removes a single edge.
\end{lemmaE}

\begin{proofE}
    Let $R_u \subseteq E$ be a subset of edges incident to $u \in V$. Assume that removing all edges in $R_u$ from $G$ reduces the total cost of $u$. Thus, we know that
    \begin{align*}
        \sum_{\{u, v\} \in R_u} \alpha \cdot w(u, v) > d_{G-R_u}(u,V) - d_{G}(u,V).
    \end{align*}
    Next, we consider the shortest paths from $u$ to all other nodes in $G$. If there is more than one shortest path to a node, we only consider one of them. Let $\{u,v\} \in R_u$ be an arbitrary edge in $R_u$. Then $Z_v \subseteq V$ denotes the set of nodes whose shortest path contains $\{u,v\}$. Note that for all edges $\{u,x_i\} \in R_u$ except for $\{u,v\}$, it holds that removing that single edge from $G$ does not increase the distance from $u$ to any node in $Z_v$. Thus, we have 
    \begin{align*}
        d_{G-R_u}(u,Z_v) - d_{G}(u,Z_v) &\geq d_{G-\{u,v\}}(u,Z_v) - d_{G}(u,Z_v) \\
        &= \sum_{\{u, x_i\} \in R_u} \left( d_{G-\{u,x_i\}}(u,Z_v) - d_{G}(u,Z_v) \right).
    \end{align*}
    Subsequently, let $Z \subseteq V$ be the set of nodes whose shortest path to $u$ contains any edge $\{u, x_i\} \in R_u$. Since the shortest path between a node $z \in Z$ and $u$ contains at most one edge $\{u, x_i\} \in R_u$, the sets $Z_{x_i}$ form a partition of $Z$. Moreover, the previous inequality holds for every $Z_{x_i}$. Thus, we can sum it up over all $Z_{x_i}$ and get
    \begin{align*}
        d_{G-R_u}(u,Z) - d_{G}(u,Z) &\geq \sum_{\{u, x_i\} \in R_u} \left( d_{G-\{u,x_i\}}(u,Z) - d_{G}(u,Z) \right).
    \end{align*}
    Due to the definition of $Z$, the distance between $u$ and nodes $v \in V \setminus Z$ does not change when removing the edges in $R_u$ from $G$. Hence, this implies that
    \begin{align*}
        d_{G-R_u}(u,V) - d_{G}(u,V) &\geq \sum_{\{u, x_i\} \in R_u} \left( d_{G-\{u,x_i\}}(u,V) - d_{G}(u,V) \right).
    \end{align*}
    Combining this with the first inequality of this proof yields
    \begin{align*}
        \sum_{\{u, v\} \in R_u} \alpha \cdot w(u, v) > \sum_{\{u, v\} \in R_u} \left( d_{G-\{u,v\}}(u,V) - d_{G}(u,V) \right)
    \end{align*}
    Therefore, since at least one term of the left sum has to be greater than its corresponding term of the right sum, there exists at least one edge whose cost is higher than the additional distance cost $u$ has to pay after the edge is removed. Hence, removing that edge is an improving move for $u$.
\end{proofE}

\begin{textAtEnd}[textEnd, category=prelim]
The next lemma provides a useful bound on the pairwise distances of nodes in any pairwise stable network $G=(V,E)$ with an arbitrary host graph $H=(V,E_H,w)$. Let $k \geq 1$. Then we call $G$ a \emph{k-spanner} if for all $u,v \in V$ we have $d_G(u,v) \leq k \cdot d_H(u,v)$. The proof of this lemma follows the proof of Bilò et al. \cite{GNCG}, who prove the same property for their model.
\end{textAtEnd}

\begin{lemmaE}[][allEnd, category=prelim]
    \label{BAE_spanner}
    Let $G=(V,E)$ be an arbitrary BGNCG network. If $G$ is pairwise stable, then $G$ is an $(\alpha + 1)$-spanner.
\end{lemmaE}

\begin{proofE}
    Let $G$ be a BGNCG network in PS and let $H=(V,E_H,w)$ be its corresponding host network. Furthermore, let $u,v \in V$ be an arbitrary pair of nodes in $G$. We split this proof into two steps. For the first step, we assume $d_H(u,v) = w(u,v)$. Subsequently, we assume towards a contradiction that $d_G(u,v) > (\alpha + 1) d_H(u,v) = (\alpha + 1)w(u,v)$. Now we consider the cost difference for $u$ and $v$ if they buy the edge $\{u,v\}$. Both nodes pay $\alpha \cdot w(u,v)$ and reduce their distance from more than $(\alpha + 1)w(u,v)$ to $w(u,v)$. This means that they make a profit of more than
    \begin{align*}
        (\alpha + 1)w(u,v) - (w(u,v) + \alpha \cdot w(u,v)) = 0.
    \end{align*}
    Thus, adding $\{u,v\}$ is an improving move for both $u$ and $v$. Therefore, $G$ is not in PS, which contradicts our assumption. Hence, $d_G(u,v) \leq (\alpha + 1) w(u,v)$.

    For the second step, we consider two arbitrary nodes $u,v \in V$ in $G$, where $d_H(u,v) \leq w(u,v)$. Let $P_{uv}=(v_0, v_1, \dotsc, v_k)$ be the shortest path between $u$ and $v$ in $H$, with $v_0=u$ and $v_k=v$. Since any subpath of a shortest path is itself a shortest path, we have $d_H(v_i, v_{i+1})=w(v_i, v_{i+1})$ for $0 \leq i \leq k-1$, that is, for any sequential pair of nodes in $P_{uv}$. Therefore, we know that
    \begin{align*}
        d_G(u,v) &\leq \sum_{i=0}^{k-1} d_G(v_i,v_{i+1}) \\ 
        &\leq \sum_{i=0}^{k-1} (\alpha + 1)d_H(v_i,v_{i+1}) \\
        &= (\alpha + 1) d_H(u,v).
    \end{align*}
    Thus, $G$ is an $(\alpha + 1)$-spanner.
\end{proofE}

\begin{textAtEnd}[textEnd, category=prelim]
The proof of the following corollary is analogous, because the existence of an improving move that only adds edges implies a possible reduction of social cost in our model.    
\end{textAtEnd}

\begin{corollaryE}[][allEnd, category=prelim]
    The social optimum network $OPT=(V,E_{OPT})$ of an arbitrary BGNCG instance is an $(\alpha + 1)$-spanner.
\end{corollaryE}

The next lemma is particularly useful for bounding distance cost. It allows us to upper bound the total distance cost of any BGNCG network, only using the distance cost of a single arbitrary node of that network. The same technique is employed for the NCG to upper bound the total cost of a network \cite{NCG_BFS_tree}. However, since this approach does not perfectly translate to a model with arbitrary edge weights, we separately bound the distance cost and the edge cost of a BGNCG network.

\begin{lemmaE}[][proofEnd, category=prelim]
    \label{min_dist_bound}
    Let $G=(V,E)$ be a BGNCG network and $z \in V$ be an arbitrary node in $G$. Then there exists a tree subgraph $T=(V, E_T)$ of $G$, with $\sum_{u \in V} d_G(u,V) \leq \sum_{u \in V} d_T(u,V) \leq 2(n-1) d_G(z,V)$.
\end{lemmaE}

\begin{proofE}
    Let $T=(V, E_T)$ be a BFS-tree of $G$ starting at $z$. We upper bound the total distance cost of $G$ by using the distances of all nodes to $z$, which are the same in $G$ and $T$ since $T$ is a BFS-tree from $z$. 
    \begin{align*}
        \sum_{u \in V} d_G(u,V) &\leq \sum_{u \in V} \sum_{v \in V} \left( d_T(u, z) + d_T(z, v) \right) \\
        &= \sum_{u \in V} \sum_{v \in V \setminus \{u\}} \left( d_G(u, z) + d_G(z, v) \right) \\
        &= \sum_{u \in V} \left( (n-1) d_G(u, z) + d_G(z, V \setminus\{u\}) \right) \\
        &= 2(n-1) d_G(z, V). \qedhere
    \end{align*}
\end{proofE}

\Cref{min_dist_bound} can now be used in combination with \Cref{remove_mult_remove_one} to obtain an upper bound on the total edge cost of any pairwise stable BGNCG network that only depends on the total distance cost of that network. This allows us to focus on bounding the total distance cost. Moreover, since networks in BNE or BSE are also pairwise stable, this holds for all our solution concepts.

\begin{theoremE}[][proofEnd, category=prelim]
    \label{edge_cost_bound_by_dist_cost}
    Let $G=(V,E)$ be a pairwise stable BGNCG network. The total edge cost in $G$ is at most $\mathcal{O} \left( \frac{\alpha}{n} \right) $ times the total distance cost in $G$, that is, $\alpha \cdot w(E) \in \mathcal{O} \left( \frac{\alpha}{n} \cdot \sum_{u \in V} d_G(u,V) \right) $.
\end{theoremE}

\begin{proofE}
    In this proof, we define a tree subgraph of $G$ and partition the edges into tree and non-tree edges. We relate the edge cost in each of these sets to the total distance cost of the subgraph, which in turn implies a bound dependent on the total distance cost of $G$.
    
    Let $v^\ast = \argmin_{u \in V}d_{G}(u,V)$ be the node with the least distance cost in $G$. We then know that the distance cost of $G$ is at least $n \cdot d_{G}(v^\ast,V)$. By \Cref{min_dist_bound}, we know that there exists a tree subgraph $T=(V,E_T)$ of $G$ with a total distance cost of at most $2(n-1) d_{G}(v^\ast,V)$. This implies that
    $$\frac{\sum_{u \in V}d_{T}(u,V)}{\sum_{u \in V}d_{G}(u,V)} \leq \frac{2(n-1)d_{G}(v^\ast,V)}{n \cdot d_{G}(v^\ast,V)} = \frac{2(n-1)}{n}.$$
    Therefore, we have
    $$\sum_{u \in V}d_{T}(u,V) \leq \frac{2(n-1)}{n} \cdot \sum_{u \in V}d_{G}(u,V) \leq 2 \cdot \sum_{u \in V}d_{G}(u,V).$$
    Since $G$ is pairwise stable and each node $u \in V$ does not delete any of its edges in $G$, we can apply \Cref{remove_mult_remove_one} and know that $u$ has at least the same cost if it removes all its edges that are not in $E_T$. As pairwise distances in $T$ are greater than or equal to the pairwise distances in $G$, we have $cost(u,G) \leq cost(u,T)$. In total, this means that
    $$\sum_{u \in V} cost(u,G) \leq \sum_{u \in V} cost(u,T).$$ This allows us to bound the edge cost of all edges $\{u, v\} \in E \setminus E_T$ by splitting the social cost of each graph into distance and edge cost. Hence, 
    $$\sum_{u \in V} d_G(u,V) + \alpha \cdot w(E_T) + \alpha \cdot w(E \setminus E_T) \leq \sum_{u \in V} d_T(u,V) + \alpha \cdot w(E_T).$$
    Rearranging for the cost of the edges $E \setminus E_T$ that are not part of $T$, we get
    $$\alpha \cdot w(E \setminus E_T) \leq \sum_{u \in V} d_T(u,V) - \sum_{u \in V} d_G(u,V).$$
    Plugging in the initial upper bound for the distance cost of $T$, we have
    \begin{align*}
        \alpha \cdot w(E \setminus E_T) &\leq 2 \cdot \sum_{u \in V}d_{G}(u,V) - \sum_{u \in V} d_G(u,V) \\
        &= \sum_{u \in V}d_{G}(u,V).
    \end{align*}
    Subsequently, we consider the edge cost of the edges $E_T$ of $T$. This is just $\alpha \cdot w(E_T)$.
    To bring this cost into relation with the total distance cost in $G$, we first evaluate how much each edge in $T$ contributes to the total distance cost in $T$. As $T$ is a tree, we know that every edge of $T$ is a bridge, i.e., removing any edge $\{u,v\} \in E_T$ results in a disconnected graph with two components $S_1,S_2 \subseteq V$. Therefore, $\{u,v\}$ is contained in at least $|S_1| \cdot |S_2| \geq (n-1)$ shortest paths between unique pairs of nodes in $T$. Thus, $\{u,v\}$ contributes at least $(n-1) \cdot w(u,v)$ to the total distance cost in $T$. Since $\sum_{u \in V}d_{T}(u,V) \leq 2 \cdot \sum_{u \in V}d_{G}(u,V)$, we have
    \begin{align*}
        \sum_{u \in V}d_{G}(u,V) &\geq \frac{1}{2} \sum_{u \in V}d_{T}(u,V) \\
        &\geq \frac{1}{2} \sum_{\{u,v\} \in E_T} (n-1) \cdot w(u,v) \\
        &\geq \frac{1}{2} (n-1) \cdot w(E_T).
    \end{align*}
    Multiplying each side of the inequality by $\frac{2\alpha}{n-1}$, we get
    \begin{align*}
        \frac{2\alpha}{n-1}\sum_{u \in V}d_{G}(u,V) &\geq \frac{2\alpha}{2(n-1)} (n-1) \cdot w(E_T) \\
        &= \alpha \cdot w(E_T).
    \end{align*}
    Finally, we sum up the edge costs of $E_T$ and $E \setminus E_T$ to calculate the total edge cost in $G$, which yields
    \begin{align*}
        \alpha \cdot w(E) &= \alpha \cdot w(E_T) + \alpha \cdot w(E \setminus E_T) \\
        &\leq \frac{2\alpha}{n-1}\sum_{u \in V}d_{G}(u,V) + \sum_{u \in V}d_{G} (u,V) 
        = \left( \frac{2\alpha}{n-1}+1 \right) \sum_{u \in V}d_{G}(u,V). \qedhere
    \end{align*}
\end{proofE}

This result is especially significant when $\alpha \in \mathcal{O}(n)$, as the total distance cost of $G$ asymptotically dominates or at least matches the total edge cost of $G$.

\section{The BGNCG on General Host Graphs}\label{sec:BGNCG}
We study the impact of cooperation on the price of anarchy if the given host graph
is an arbitrarily positively weighted graph. We show the negative result that the PoA in the BGNCG is asymptotically the same, and very high, for all solution concepts. Thus, in general increased cooperation on its own does not yield better networks. 

The following theorem introduces a simple construction for a lower bound on the PoA for BGNCG networks in BSE. As the proposed network is in BSE, this bound holds for all our solution concepts.

\begin{theoremE}[][proofEnd, category=BGNCG]
    \label{PoA_lower_BGNCG_BSE_alpha}
    For $n > 2$, the PoA for BGNCG networks in BSE is at least $\alpha + 1$.
\end{theoremE}

\begin{figure}[ht]
        \centering
        \begin{subfigure}[t]{0.25\linewidth}
            \includegraphics[width=\linewidth]{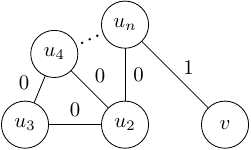}
            \label{fig:BGNCG_BSE_OPT}
        \end{subfigure}
        \hskip 4em
        \begin{subfigure}[t]{0.25\linewidth}
            \includegraphics[width=\linewidth]{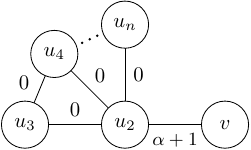}
            \label{fig:BGNCG_BSE}
        \end{subfigure}
        \caption{Lower bound construction for the PoA of BGNCG networks in BSE. Left: BGNCG network $G_n^\ast$ with a low social cost. Right: BGNCG network $G_n$ in BSE. For any edge $\{u_i,u_j\}$, it holds that $w(u_i,u_j) = 0$. The remaining edges that are not depicted have weight $1$.}
        \label{fig:BGNCG_BSE_lower_bound}
\end{figure}

\begin{proofE}
    The social cost ratio between $G_n$ and $G_n^\ast$ in \Cref{fig:BGNCG_BSE_lower_bound} is
    $$\frac{cost(G_n)}{cost(G_n^\ast)} \leq \frac{(2n-2+2\alpha)(\alpha + 1)}{(2n-2+2\alpha)1}=\alpha + 1.$$
    We claim that the network $G_n$ is in BSE. Only removing edges is not an improving move for any coalition of nodes, since it would disconnect the graph or not reduce cost at all. Therefore, any coalition can only improve by buying at least one edge $\{u_i, v\}$ of weight $1$. This edge costs $\alpha$ and decreases the cost of $u_i$ by at most $\alpha$. Thus, $u_i$ does not improve and there exists no coalition of nodes with an improving move, that is, $G_n$ is in BSE.
\end{proofE}

Now we provide an almost matching upper bound on the PoA of pairwise stable BGNCG networks. The proof strategy is the same as the one Friedemann, Friedrich, Gawendowicz, Lenzner, Melnichenko, Peters, Stephan, and Vaichenker~\cite{Efficiency_and_Stability_in_Euclidean_Network_Design} use to prove a similar bound for the GNCG.

\begin{theoremE}[][proofEnd, category=BGNCG]
    \label{PoA_upper_BGNCG_PS_alpha}
    The Price of Anarchy for pairwise stable BGNCG networks is at most $2(\alpha + 1)$.
\end{theoremE}

\begin{proofE}
    Let $G=(V,E)$ be BGNCG network in PS and let $OPT=(V,E_{OPT})$ be the social optimum network. We bound the total distance cost of $G$ and total edge cost of $G$ separately. 
    
    To bound the total distance cost we use \Cref{BAE_spanner}. This immediately yields the following inequality:
    $$\sum_{u,v \in V} d_G(u,v) \leq \sum_{u,v \in V} (\alpha + 1) d_H(u,v) \leq \sum_{u,v \in V} (\alpha + 1) d_{OPT}(u,v).$$
    In order to analyze the edge cost of $G$, we partition the edges into two sets. For any two nodes $u,v \in V$, let $\pi_G(u,v) \subseteq E$ be a shortest path from $u$ to $v$ in $G$. Subsequently, the partition is defined by the set $P \coloneqq \{\pi_G(u,v) \mid \{u,v\} \in E_{OPT}\}$. For every edge $\{u,v\}$ in $OPT$ the set $P$ contains the edges of the shortest $uv$-path in $G$. We denote the other set of the partition as $R \coloneqq E \setminus P$. 
    
    Again using \Cref{BAE_spanner} and by definition of $P$, we bound the edge cost of $P$ as follows:
    \begin{align*}
        \alpha \cdot w(P) &\leq \alpha \smashoperator\sum_{\{u,v\} \in E_{OPT}} d_G(u,v) \leq \alpha (\alpha + 1) \sum_{\{u,v\} \in E_{OPT}} d_H(u,v) \\
        &\leq \alpha (\alpha + 1) \cdot w(E_{OPT}).
    \end{align*}
    
    Next, we evaluate the edge cost of $R$ by considering the cost change of an arbitrary node $z \in V$ if it removes all of its edges in $R$. We denote that set of edges as $R_z$. Since $G$ is in PS, we know that
    \begin{align*}
        d_G(z, V) + \alpha \cdot w(R_z) \leq d_{G-R_z}(z, V),
    \end{align*}
    where $G-R_z$ is the network obtained after removing all edges of $R_z$ from $G$. Therefore, we have 
    \begin{align*}
        \alpha \cdot w(R_z) \leq d_{G-R_z}(z, V) - d_G(z, V) \leq d_{G-R_z}(z, V).
    \end{align*}
    Now, we further analyze the distance between $z$ and another arbitrary node $x \in V$ in $G-R_z$. We do this by considering their shortest path $\pi_{OPT}(z,x) = (v_0, v_1, \dotsc, v_k)$ in $OPT$. For every edge $\{v_i, v_{i+1}\}$ on $\pi_{OPT}(z,x)$, by definition of $P$, there exists a shortest path $\pi_{G-R_z}(v_i,v_{i+1})$ in $G-R_z$ that has a length of at most $(\alpha + 1) d_H(v_i,v_{i+1}) \leq (\alpha + 1) d_{OPT}(v_i,v_{i+1})$. Hence, the following holds:
    \begin{align*}
        d_{G-R_z}(z, x) &\leq \sum_{i=0}^{k-1} d_{G-R_z}(v_i, v_{i+1}) \\
        &\leq \sum_{i=0}^{k-1} (\alpha + 1) d_{OPT}(v_i,v_{i+1}) = (\alpha + 1) d_{OPT}(z,x).
    \end{align*}
    As this holds for any two arbitrary nodes, we conclude that
    $$ 2\alpha \cdot w(R) \leq \sum_{z \in V} d_{G-R_z}(z,V) \leq \sum_{z \in V} (\alpha + 1) d_{OPT}(z,V). $$
    Combining this with the edge cost bound of $P$ and the distance cost bound we get a total social cost ratio of
    \begin{align*}
        \frac{cost(G)}{cost(OPT)} &= \frac{2\alpha \cdot w(P) + 2\alpha \cdot w(R) + \sum_{u \in V} d_G(u,V)}{2\alpha \cdot w(E_{OPT}) + \sum_{u \in V} d_{OPT}(u,V)} \\
        &\leq \frac{2\alpha (\alpha+1) \cdot w(E_{OPT}) + 2(\alpha + 1) \sum_{u \in V} d_{OPT}(u,V)}{2\alpha \cdot w(E_{OPT}) + \sum_{u \in V} d_{OPT}(u,V)} \\
        &\leq \frac{2(\alpha+1) \left( 2\alpha \cdot w(E_{OPT}) + \sum_{u \in V} d_{OPT}(u,V) \right)}{2\alpha \cdot w(E_{OPT}) + \sum_{u \in V} d_{OPT}(u,V)} \\
        &= 2(\alpha + 1). \qedhere
    \end{align*}
    
\end{proofE}

\section{The BGNCG on Host Graphs with Metric Edge Weights}\label{sec:metric}

This section contains our results on the natural BGNCG special case where the given host graph has metric edge weights, e.g., these could be Euclidean distances between points in $\mathbb{R}^2$, which models the case of the decentralized construction of a fiber-optic communication network between cities. 

We provide several lower bounds and upper bounds on the price of anarchy for M-BGNCG networks. As our main result, we provide an upper bound that is asymptotically tight for $\alpha\in \mathcal{O}(n)$. Along the way, we also provide a tight bound for $\alpha \in \Omega(n^2)$. 

Our starting point are constructions of networks that yield non-trivial lower bounds on the price of anarchy for M-BGNCG networks in PS, BNE, and BSE.

\begin{figure}[ht]
    \centering
    \begin{subfigure}[t]{0.22\linewidth}
        \includegraphics[width=0.95\linewidth]{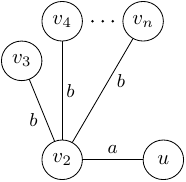}
        \caption{The network $S_n^\ast$ with low social cost.} \label{fig:NE_general_OPT}
    \end{subfigure}
    \hskip 1em
    \begin{subfigure}[t]{0.23\linewidth}
        \includegraphics[width=0.95\linewidth]{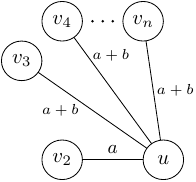}
        \caption{The stable network $S_n$.} \label{fig:NE_general}
    \end{subfigure}
    \hskip 1em
    \begin{subfigure}[t]{0.23\linewidth}
        \includegraphics[width=0.95\linewidth]{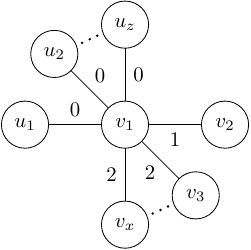}
        \caption{The network $G_n^\ast$ with a low social cost.} \label{fig:M-BGNCG_BSE_dist_OPT}
    \end{subfigure}
    \hskip 1em
    \begin{subfigure}[t]{0.24\linewidth}
        \includegraphics[width=0.95\linewidth]{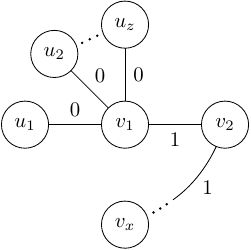}
        \caption{The network $G_n$ in BSE.} \label{fig:M-BGNCG_BSE_dist_BSE}
    \end{subfigure}
    \caption{(a) and (b): lower bound constructions for the PoA for M-BGNCG networks in PS, BNE, and BSE. The values of $a$ and $b$ differ depending on the solution concept used. The edge weight of an arbitrary edge $\{u,v\}$ of the host graph of $S_n^\ast$ and $S_n$ is defined as $d_{S_n^\ast}(u,v)$. (c) and (d): another lower bound construction for the PoA for M-BGNCG networks in BSE. For both graphs holds that $x = \lfloor \frac{\sqrt{\alpha}}{2} \rfloor$ and $z = n - \lfloor \frac{\sqrt{\alpha}}{2} \rfloor$. For any edge $\{v_i,v_j\}$, it holds that $w(v_i,v_j) = 1$, if $|j-i|=1$, and $w(v_i,v_j) = 2$ otherwise. All other edge weights can be inferred from the triangle inequality.}
    \label{fig:M-BGNCG_lower_bounds}
\end{figure}

\begin{theoremE}[][proofEnd, category=metric]
\label{M-BGNCG_PoA_lower_bounds}
The PoA for M-BGNCG networks is in
\begin{enumerate}
\item[(1)] $\Omega(\min\{n, \alpha \})$ for PS,
\item[(2)] $\Omega(\min\{n, \sqrt{\alpha}\})$ for BNE,
\item[(3)] $\Omega(\min\{n, \frac{\alpha}{n}\})$ for BSE,
\item[(4)] $\Omega \left( \frac{n\sqrt{\alpha}}{n+\alpha} \right)$ for BSE, if $\alpha\le n^2$.
\end{enumerate}
\end{theoremE}

\begin{proofE}
    We show the statements (1-3) with the same generic instance, depicted in \Cref{fig:NE_general_OPT} and \Cref{fig:NE_general}, that we instantiate with different values for the edge weights $a$ and $b$. Statement (4) is shown via the instance in \Cref{fig:M-BGNCG_BSE_dist_OPT} and \Cref{fig:M-BGNCG_BSE_dist_BSE}. We assume that $\alpha \in \omega(1)$ since the theorem trivially holds for $\alpha \in \mathcal{O}(1)$, as the PoA is by definition in $\Omega(1)$ if an equilibrium exists. In the following, we commonly use the formula $(2n - 2 + 2\alpha) \cdot w(E_S)$ for the social cost of a star graph $S$.

    \paragraph{Proof of (1):}
    For $a=1$ and $b= \frac{2}{\alpha}$, the graph $S_n^\ast=(V_n,E_{S_n^\ast})$ in \Cref{fig:NE_general_OPT} has a social cost of
    \begin{align*}
        cost(S_n^\ast) &= (2n - 2 + 2\alpha) \cdot w(E_{S_n^\ast}) \\
        &= (2n - 2 + 2\alpha) \cdot ((n-2)\frac{2}{\alpha} + 1).
    \end{align*}
    Now we consider the network $S_n=(V_n, E_n)$ in \Cref{fig:NE_general}. We claim that it is in PS. As removing any edge disconnects the network, there exists no improving move that removes an edge. The center agent~$u$ is not able to buy any edges. Thus, it suffices to show that no leaf agent~$v_i$ can improve by buying an edge. Buying an additional edge costs at least $\alpha \cdot \frac{2}{\alpha}=2$ for any leaf agent and yields a distance benefit of exactly $2$ for any edge. Hence, there does not exist an improving move for any agent and $S_n$ is pairwise stable.
    
    For $a=1$ and $b= \frac{2}{\alpha}$, the social cost of $S_n$ is
    \begin{align*}
        cost(S_n) &= (2n - 2 + 2\alpha) \cdot w(E_n) \\
        &= (2n - 2 + 2\alpha) \cdot ((n-2) (\frac{2}{\alpha} + 1) + 1).
    \end{align*}
    Hence, the social cost ratio between the pairwise stable network $S_n$ and the network $S_n^\ast$ is
    \begin{align*}
        \frac{cost(S_n)}{cost(S_n^\ast)} &= \frac{(n-2) (\frac{2}{\alpha} + 1) + 1}{(n-2)\frac{2}{\alpha} + 1} \\
        &= 1 + \frac{n-2}{(n-2)\frac{2}{\alpha} + 1}.
    \end{align*}
    Thus, the PoA for M-BGNCG networks in PS is in $\Omega(1 + \frac{n}{n / \alpha + 1})$, which is constant if $\alpha \in \mathcal{O}(1)$. For $\alpha \in \Theta(n^z)$ with $z \in \R_0^+$, we have
    \begin{align*}
                \Omega(1+\frac{n}{n/\alpha + 1}) &=
        \begin{cases}
            \Omega(n^z) = \Omega(\alpha), &\text{if } 0 \leq z \leq 1; \\
            \Omega(n), &\text{else.}
        \end{cases}
    \end{align*}

    \paragraph{Proof of (2):}
    For $a=1$ and $b=\frac{2}{\sqrt{\alpha}}$, the graph $S_n^\ast=(V_n,E_{S_n^\ast})$ in \Cref{fig:NE_general_OPT} has a social cost of
    \begin{align*}
        cost(S_n^\ast) =& (2n - 2 + 2\alpha) \cdot w(E_{S_n^\ast}) \\
                  =& (2n - 2 + 2\alpha) \cdot ((n-2)\frac{2}{\sqrt{\alpha}} + 1).
    \end{align*}
    Network $S_n=(V_n,E_n)$ in \Cref{fig:NE_general} is in BNE. To verify this, we consider the possible strategy changes for all agents. The center agent~$u$ cannot improve since its only move is to delete a set of bridges. Moreover, note that, for $\alpha \geq 1$, only buying edges is not an improving move for any agent~$v_i$ as every edge costs at least $2\sqrt{\alpha}$ and only reduces distance cost by $2$. Hence, the only potentially improving move is for a leaf agent~$v_i$ to remove its edge and build multiple edges to other leaves. Deleting an edge reduces the cost by at most $\alpha + 2\sqrt{\alpha}$, which allows agent~$v_i$ to buy at most $\frac{\sqrt{\alpha}}{2}+1$ edges while still making profit. Let $v_j$ be a node agent~$v_i$ builds an edge towards. As agent~$v_j$ pays at least $2\sqrt{\alpha}$ and every other edge agent~$v_i$ builds reduces the distance cost of agent~$v_j$ by at most $2$, the number of edges agent~$v_i$ needs to buy in order for agent~$v_j$ to accept the change has to be at least $\sqrt{\alpha}$. This exceeds the amount of edges agent~$v_j$ can buy without increasing its own cost. Hence, there does not exist any agent~$v \in V$ in $S_n$ with an improving move with respect to BNE. Thus, $S_n$ is in BNE.
    For $a=1$ and $b=\frac{2}{\sqrt{\alpha}}$, the social cost of $S_n$ is
    \begin{align*}
        cost(S_n) =& (2n - 2 + 2\alpha) \cdot w(E_n) \\
                       =& (2n - 2 + 2\alpha) \cdot ((n-2) (\frac{2}{\sqrt{\alpha}} + 1) + 1).
    \end{align*}
    Thus, the social cost ratio is
    \begin{align*}
        \frac{cost(S_n)}{cost(S_n^\ast)} =& \frac{(n-2) (\frac{2}{\sqrt{\alpha}} + 1) + 1}{(n-2)\frac{2}{\sqrt{\alpha}} + 1} \\
                                    =& 1 + \frac{(n-2)}{(n-2)\frac{2}{\sqrt{\alpha}} + 1}.
    \end{align*}
    So the social cost ratio is in $\Omega(1+\frac{n}{n/\sqrt{\alpha} + 1})$, which is constant if $\alpha \in \mathcal{O}(1)$. For $\alpha \in \Theta(n^z)$ with $z \in \R_0^+$, we have
    \begin{align*}   
    \Omega(1+\frac{n}{n/\sqrt{\alpha} + 1}) &=
    \begin{cases}
        \Omega(\sqrt{n^z}) = \Omega(\sqrt{\alpha}), &\text{if } 0 \leq z \leq 2; \\
        \Omega(n), &\text{else.}
    \end{cases}
    \end{align*}

    \paragraph{Proof of (3):}
    For $a= \frac{1}{n}$ and $b=\frac{2}{\alpha}$, the graph $S_n^\ast=(V_n, E_{S_n^\ast})$ in \Cref{fig:NE_general_OPT} has a social cost of
    \begin{align*}
        cost(S_n^\ast) =& (2n - 2 + 2\alpha) \cdot w(E_{S_n^\ast}) \\
                  =& (2n - 2 + 2\alpha) \cdot ((n-2)\frac{2}{\alpha} + \frac{1}{n}).
    \end{align*}
    
    Network $S_n=(V_n, E_n)$ in \Cref{fig:NE_general} is in BSE. To verify this, we show that there exists no coalition of agents $\Gamma \subseteq V_n$ with an improving move with respect to BSE.
    The center agent~$u$ does not have to be considered for coalitions as it can only delete edges that are bridges, which can be done unilaterally. Thus, any coalition with an improving move that includes agent~$u$ is still a coalition with an improving move if agent~$u$ is removed from the coalition. Hence, we only need to consider coalitions of leave agents~$v_i$. As edges can only be built between members of the coalition, there has to be at least one agent~$v_i$ in any coalition that keeps its edge to node $u$ in order to keep the network connected. As agent~$v_i$ does not remove any edges, it can only improve its cost by buying at least one edge. Any edge agent~$v_i$ can buy costs at least $2$ and even if buying that edge would minimize agent~$v_i$'s distance to all other nodes, its distance cost would only improve by $(n-2) \cdot \frac{2}{n} < 2$. Hence, $S_n$ is in BSE.
    
    For $a= \frac{1}{n}$ and $b=\frac{2}{\alpha}$, the social cost of the BSE $S_n$ is
    \begin{align*}
        cost(S_n) =& (2n - 2 + 2\alpha) \cdot w(E_n) \\
                       =& (2n - 2 + 2\alpha) \cdot ((n-2) (\frac{2}{\alpha} + \frac{1}{n}) + \frac{1}{n}).
    \end{align*}
    Thus, the social cost ratio is
    \begin{align*}
        \frac{cost(S_n)}{cost(S_n^\ast)} =& \frac{(n-2) (\frac{2}{\alpha} + \frac{1}{n}) + \frac{1}{n}}{(n-2)\frac{2}{\alpha} + \frac{1}{n}} \\
                                         =& 1 + \frac{(n-2)\frac{1}{n}}{(n-2)\frac{2}{\alpha} + \frac{1}{n}} \\
                                         =& 1 + \frac{n-2}{\frac{2n(n-2)}{\alpha} + 1}.
    \end{align*}
    So the social cost ratio is in $\Omega(1+\frac{n}{n^2/\alpha + 1})$, which is constant if $\alpha \in \mathcal{O}(n)$. For $\alpha \in \Theta(n^{1+z})$ with $z \in \R_0^+$, we have
    \begin{align*}
    \Omega(1+\frac{n}{n^2/\alpha + 1}) &=
    \begin{cases}
        \Omega(n^z) = \Omega(\frac{\alpha}{n}), &\text{if } 0 \leq z \leq 1; \\
        \Omega(n), &\text{else.}
    \end{cases}
    \end{align*}

    \paragraph{Proof of (4):}
    We assume that $\alpha \in \omega(1)$.
    There exists a constant $c_1 \in \R^+$, so that the graph $G_n^\ast=(V_n, E_{G_n^\ast})$ in \Cref{fig:M-BGNCG_BSE_dist_OPT} has a social cost of
    \begin{align*}
        cost(G_n^\ast) &\leq c_1 \cdot \left( \alpha \sqrt{\alpha} + n\sqrt{\alpha} \right) \\
        &= c_1 \cdot \sqrt{\alpha}(n + \alpha).
    \end{align*}
    
    \Cref{fig:M-BGNCG_BSE_dist_BSE} shows the network $G_n=(V_n,E_n)$ in BSE. We prove it is in fact in BSE by showing that in any coalition $\Gamma \in V_n$ with some strategy change, the agent with the least distance to $v_1$ in $G_n$ does not improve.

    Let $u \in V_n$ be an arbitrary node in $G_n$. Moreover, let $R(u)$ be defined as the set of nodes that have a greater distance to node~$v_1$ in $G_n$ than node $u$ has.
    First, we consider the cost of one of the agents~$u_i \in V_n$ in $G_n$, with $d_{G_n}(u_i,v_1) = 0$, that only own edges of weight $0$. Since the longest path in $G_n$ has a length of less than $\sqrt{\alpha} \cdot 1$, and agent~$u_i$ only has distance cost $R(u_i)$, it has a total cost of less than
    $$|R(u_i)| \cdot \sqrt{\alpha} < \sqrt{\alpha} \cdot \sqrt{\alpha} = \alpha.$$
    Furthermore, any other strategy that includes an edge of non-zero weight would incur a cost of at least $\alpha$ which is more than the cost of $u_i$ in $G_n$. Among the remaining strategies that only use edges of weight~$0$, all strategies that connect agent~$u_i$ to the network have the same cost, which is why agent~$u_i$ does not change its strategy as long as there does not exist a coalition without agent~$u_i$ that has an improving move. Therefore, from now on, we ignore any coalitions containing an agent~$u_i$ that only owns edges of weight~$0$ in $G_n$.

    Consider an arbitrary coalition $\Gamma \in V_n$ with some strategy change. Let $G_n'$ be the graph after performing that change. Let $v_i \in V_n$ be the node in that coalition with the least distance to node~$v_1$ in $G_n$. As no edge between $\Gamma$ and $V_n \setminus \Gamma$ is built in $G_n'$, the subgraph induced by $V_n \setminus R(v_i)$ is still a tree in $G_n'$. This is the case because by changing from $G_n$ to $G_n'$ the agent~$v_i$ does not remove any edges to nodes in $V_n \setminus \Gamma$, as removing these edges would disconnect the graph. Furthermore, this implies that modifying $G_n$ to $G_n'$ does not decrease agent~$v_i$'s distance to $V_n \setminus \Gamma$, but rather only potentially to $R(v_i)$. Hence, if $R(v_i) = \emptyset$, then agent~$v_i$ does not improve. Subsequently, we consider the case $R(v_i) \neq \emptyset$. The longest path in $G_n$ has a length of less than $\sqrt{\alpha}$ and therefore the distance cost of agent~$v_i$ to $R(v_i)$ in $G_n$ is $d_{G_n} \left( v_i, R(v_i) \right) < \sqrt{\alpha}\sqrt{\alpha}=\alpha$. Since agent~$v_i$ owns exactly one edge to $R(v_i)$ in $G_n$, which has a cost of $\alpha$, and every other possible set of edges between agent~$v_i$ and nodes in $R(v_i)$ contains an edge that costs at least $2\alpha$, the cost benefit for agent~$v_i$ if it changes its set of edges is at most
    \begin{align*}
        cost(v_i,G_n) - cost(v_i,G_n') &\leq d_{G_n}(v_i, R(v_i)) - (d_{G_n'}(v_i, R(v_i)) + \alpha) \\
        &< \alpha - 0 - \alpha \\
        &=0.
    \end{align*}
    Consequently, agent~$v_i$ does not change its strategy and either $\Gamma$ is not a coalition with an improving move or there exists a coalition $\Gamma' \in \Gamma \setminus \{v_i\}$ with an improving move. However, this holds for any agent~$v_i \in V_n$ that owns at least one edge of non-zero weight in $G_n$. Thus, for any agent~$u \in V_n$, there exists no coalition $\Gamma \in V_n$ with an improving move where $u \in \Gamma$ holds, that is, there exists no coalition $\Gamma \in V_n$ with an improving move. Hence, $G_n$ is in BSE.
    
    There exists a constant $c_2 \in \R^+$, so that the BSE $G_n$ has a social cost of
    \begin{align*}
        cost(G_n) &\geq c_2 \cdot \left( \alpha \sqrt{\alpha} + n\sqrt{\alpha}\sqrt{\alpha} \right)
        \geq c_2 \cdot n\alpha.
    \end{align*}
    Therefore, there exists a constant $c \in \R^+$, such that the social cost ratio of $G_n$ and $G_n^\ast$ is at least
    \begin{align*}
        \frac{cost(G_n)}{cost(G_n^\ast)} \geq c \cdot \frac{n\alpha}{\sqrt{\alpha}(n+\alpha)} &= c \cdot \frac{n\sqrt{\alpha}}{n+\alpha}.\qedhere
    \end{align*}
\end{proofE}

For $\alpha \in \mathcal{O}(n)$, the PoA lower bound obtained by the instance in~\Cref{fig:M-BGNCG_BSE_dist_OPT} and \Cref{fig:M-BGNCG_BSE_dist_BSE} is especially high, that is, in $\Omega \left( \sqrt{\alpha} \right)$. This is because, unlike the PoA lower bound construction in \Cref{fig:NE_general_OPT} and \Cref{fig:NE_general}, this lower bound construction heavily depends on a high distance cost ratio, which is why it is higher relative to $\alpha$ for smaller values of $\alpha$.

The following theorem provides an upper bound on the price of anarchy for pairwise stable M-BGNCG networks. Since networks in BNE or BSE are also pairwise stable, this bound applies to these solution concepts as well. However, we later show better upper bounds for networks in BSE. The proof of the following theorem is close to the proof by Bilò, Friedrich, Lenzner, and Melnichenko~\cite{GNCG}, who prove a similar bound for GNCG networks in Nash equilibrium.

\begin{theoremE}[][proofEnd, category=metric]
    \label{PoA_upper_M-BGNCG_PS_alpha}
    The PoA for pairwise stable M-BGNCG networks is at most $\alpha + 1$.
\end{theoremE}

\begin{proofE}
    Let $G=(V,E)$ be a pairwise stable M-BGNCG network and let $u, v \in V$ be two distinct nodes. Let $x$ and $x^\ast$ be two indicator variables such that $x = 1$ iff $\{u, v\}$ is an edge of $G$ and $x^\ast  = 1$ iff $\{u, v\}$ is an edge of the social optimum network $OPT=(V,E_{OPT})$. We prove the claim by showing that
    \begin{align*}
        \sigma := \frac{2\alpha \cdot w(u,v) \cdot x + 2d_G(u,v)}{2\alpha \cdot w(u,v) \cdot x^\ast + 2d_{OPT}(u,v)} \leq \alpha + 1.
    \end{align*}
    Essentially, $\sigma$ is the ratio of the social cost contribution of a pair of nodes between the pairwise stable network $G$ and social optimum network $OPT$. If this ratio is bounded by $\alpha + 1$ for every pair of nodes, this also holds for their sum. The proof is split into two cases. If $x = 1$, we have $d_G(u,v) = w(u,v)$ and therefore
    \begin{align*}
        \sigma \leq \frac{2(\alpha + 1) w(u,v)}{2d_{OPT}(u,v)} \leq \frac{2(\alpha + 1) w(u,v)}{2w(u,v)} = \alpha + 1.
    \end{align*}
    If $x = 0$, then it holds by \Cref{BAE_spanner} that
    \begin{align*}
        \sigma \leq \frac{2d_G(u,v)}{2d_{OPT}(u,v)} &\leq \frac{(\alpha + 1)d_H(u,v)}{d_H(u,v)} = \alpha + 1. \qedhere
    \end{align*}
    \begin{figure}[b]
        \centering
        \begin{subfigure}{0.4\linewidth}
            \includegraphics[width=0.7\linewidth]{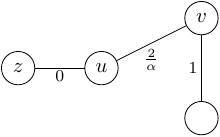}
            \caption{The host network $H=(V, E_H,w)$.} \label{fig:sigma_worst_case_host}
        \end{subfigure}
        \hskip 5em
        \begin{subfigure}{0.4\linewidth}
            \includegraphics[width=0.7\linewidth]{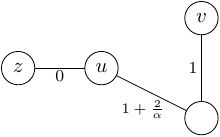}
            \caption{The pairwise stable network $G=(V,E)$.} \label{fig:sigma_worst_case}
        \end{subfigure}
        \caption{A M-BGNCG instance containing a pair of nodes $u,v \in V$ with value of $\sigma$ of $\alpha + 1$.}
        \label{fig:M-BGNCG_PS_upper_bound_sigma_worst_case}
    \end{figure}
\end{proofE}

\begin{textAtEnd}[textEnd, category=metric]
In contrast to the bounds on the PoA for pairwise stable GNCG networks, this leaves us with an upper bound that is not equal to the lower bound. However, it is the best upper bound this proof strategy can yield. To see this, consider the M-BGNCG instance in \Cref{fig:M-BGNCG_PS_upper_bound_sigma_worst_case}. As the network $G$ in \Cref{fig:sigma_worst_case} is a tree, no agent can benefit from removing an edge. Furthermore, the only edges missing in $G$ that would decrease distance cost are $\{u,v\}$ and $\{z,v\}$, and adding one of those edges costs $\frac{2}{\alpha} \cdot \alpha = 2$ and reduces the distance of agent~$u$ or agent~$z$, respectively, by $2 + \frac{2}{\alpha} - \frac{2}{\alpha} = 2$.  Therefore, network~$G$ is pairwise stable. For agents~$u$ and $v$ the definition of $\sigma$ yields a value of $\frac{2d_G(u,v)}{2d_{OPT}(u,v)} = \frac{2 + \frac{2}{\alpha}}{\frac{2}{\alpha}} = \alpha + 1$. Thus, no lower value for $\sigma$ can be proven for pairwise stable M-BGNCG networks.
\end{textAtEnd}

In the NCG and many of its variants, the PoA is often bounded dependent on $\alpha$. However, this does not always give the full picture. In the BGNCG, the instance parameter $n$ also plays an important role with respect to the price of anarchy. The next lemmas and theorems analyze distance cost ratios, edge cost ratios, and the PoA with respect to the number of agents~$n$. 

The following lemma bounds the distance cost ratio between any connected M-BGNCG network and its corresponding social optimum network. Note, that for this, the M-BGNCG network does not even have to be in any kind of equilibrium.

\begin{lemmaE}[][normal, category=metric]
    In the M-BGNCG, the distance cost ratio between any connected subgraph $G=(V,E)$ of $H=(V,E_H,w)$ and $OPT=(V,E_{OPT})$ is at most $2(n-1)$. \label{lem:path_length}
\end{lemmaE}

\begin{proofE}
    Let $v_0, v_k \in V$ be two arbitrary nodes and let $P_{v_0v_k} = (v_0, v_1, ..., v_{k-1}, v_k)$ be the shortest path between these nodes. Using the triangle inequality we can bound the length of this path as follows:
    \begin{align*}
        d_G(v_0, v_k) &= \sum_{i=0}^{k-1} w(v_i, v_{i+1})\leq w(v_0, v_1) + \sum_{i=1}^{k-1} \left( w(v_0, v_i) + w(v_0, v_{i+1}) \right)\leq 2 \sum_{i=1}^k w(v_0, v_i).
    \end{align*}
    Since $v_0$ and $v_k$ are arbitrarily chosen, we have for all $u \in V$ that
    \begin{align*}
        d_G(u, V) \leq 2(n-1) \sum_{v \in V \setminus \{u\}} w(u, v).
    \end{align*}
    If we sum this up over all nodes, we count each edge at most $4(n-1)$ times. Since every edge is the shortest path between the nodes it connects, the optimum distance cost is at least twice the sum of all edge weights. This results in a distance cost ratio of at most $2(n-1)$.
\end{proofE}

Subsequently, we upper bound the edge cost ratio between any M-BGNCG network that is a tree and its corresponding social optimum network. We get the following bound:

\begin{lemmaE}[][normal, category=metric]
    In the M-BGNCG, the edge cost ratio between any tree subgraph $T=(V,E_T)$ of the host graph $H$ and the social optimum network $OPT=(V,E_{OPT})$ is at most $n$.
\end{lemmaE}

\begin{proofE}
    Let $MinST=(V,E_{Min})$ be the minimum spanning tree of the host graph $H=(V,E_H,w)$ and let $MaxST=(V,E_{Max})$ be the maximum spanning tree of $H$. Moreover, let $\{v_0, v_1\} = \argmax_{\{u, v\} \in E_H} w(u,v)$. Since the triangle inequality holds, we know that $d_{MinST}(v_0, v_1) \geq w(v_0, v_1)$. Thus, we have:
    \begin{align*}
        \frac{2\alpha \cdot w(E_T)}{2\alpha \cdot w(E_{OPT})} &\leq \frac{w(E_{Max})}{w(E_{Min})}\leq \frac{n \cdot w(v_0, v_1)}{w(v_0, v_1)}= n.
    \end{align*}
    This directly yields the desired upper bound for the edge cost ratio between any tree $T$ and $OPT$.
\end{proofE}

This already allows us to upper bound the PoA restricted to M-BGCNG networks that are trees. Since the last two lemmas only use the property of metric edge weights, this upper bound holds for any kind of solution concept. Thus, we have:

\begin{corollaryE}[][normal, category=metric]
\label{M-BGNCG_tree_PoA}
    In the M-BGNCG, the PoA for pairwise stable tree networks is at most $2(n-1)$.
\end{corollaryE}

We now use \Cref{M-BGNCG_tree_PoA} and \Cref{remove_mult_remove_one} to show the same PoA upper bound for any pairwise stable M-BGNCG network.

\begin{theoremE}[][normal, category=metric]
    \label{PoA_upper_M-BGNCG_PS_n}
    In the M-BGNCG, the PoA for pairwise stable networks is at most $2(n-1)$.
\end{theoremE}

\begin{proofE}
    Let $G=(V,E)$ be a pairwise stable network and let $T=(V,E_T)$ be a tree subgraph of $G$. Such a tree exists since $G$ is pairwise stable and therefore connected. Now, we consider an arbitrary agent~$u \in V$. As agent~$u$ does not change its strategy and by \Cref{remove_mult_remove_one}, we know that agent~$u$'s total cost would be at least as great as in $G$ if it only bought the edges it buys in $T$. Furthermore, every pairwise distance in $G$ is at most as great as it is in $T$. Thus, it follows that $cost(u, G) \leq cost(u, T)$. 
    
    Since this holds for all agents, we have
    $cost(G) \leq cost(T)$.
    This allows us to use \Cref{M-BGNCG_tree_PoA}, as $T$ is a tree and the social cost of $G$ is less than or equal to the social cost of $T$.
\end{proofE}

Combining \Cref{PoA_upper_M-BGNCG_PS_n} with \Cref{M-BGNCG_PoA_lower_bounds} results in an asymptotically tight bound on the PoA for M-BGNCG networks in PS, BNE, and BSE, for $\alpha \in \Omega(n^2)$.

\begin{corollaryE}[][normal, category=metric]
    For $\alpha \in \Omega(n^2)$, the PoA in the M-BGNCG is in $\Theta \left( n \right)$ for PS, BNE, and BSE.
\end{corollaryE}

The following theorem is the key technical result of this paper. It bounds the distance cost ratio between M-BGNCG networks in BSE and their corresponding social optimum networks.

\begin{theoremE}[][proofEnd, category=metric]
    \label{M-BGNCG_dist_cost_ratio_BSE}
    In the M-BGNCG, for $\alpha \in \omega(1)$ and $\alpha \leq n^2$ the distance cost ratio between any network $G = (V,E)$ in BSE and the social optimum network $OPT=(V, E_{OPT})$ is in $\mathcal{O}(\sqrt{\alpha})$.
\end{theoremE}

\begin{proofE}
    \begin{figure}[h]
        \centering
        \includegraphics[width=\linewidth]{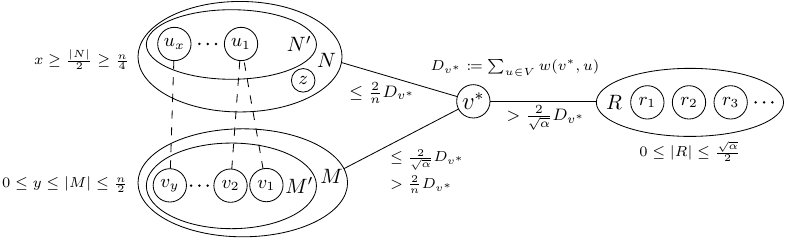}
        \caption{The general structure of the proof of \Cref{M-BGNCG_dist_cost_ratio_BSE}. The edges between $v^\ast$ and the three sets show edge weight bounds for the edges between $v^\ast$ and any node of that set, respectively. The dashed lines represent the improving move in $G$. Every node $v_i \in M'$ buys exactly one edge to a node $u_i \in N'$, so that every node in $N'$ has to buy at most two edges.}
        \label{fig:M-BGNCG_dist_ratio_BSE_proof_sketch}
    \end{figure}
    The idea of the proof is to sort and partition all agents by the weight of their edge to one specific node. Due to the metric property of the edge weights, this allows us to easier bound the distance cost of agents from different partitions. We then show that there exists an improving move if the distance cost of each agent exceeds a certain threshold. This improving move is formed by two sets of agents from different partitions. Each agent in one set buys exactly one edge to a node in the other set of size $\Omega(n)$ such that all agents buy a constant number of edges. \Cref{fig:M-BGNCG_dist_ratio_BSE_proof_sketch} depicts the general structure of the idea of this proof.

    Let $v^\ast = \argmin_{u \in V}d_{OPT}(u,V)$ be the agent with the least distance cost in $OPT$. We then know that the distance cost of $OPT$ is at least $n \cdot d_{OPT}(v^\ast,V)$. Since we have metric edge weights, the triangle inequality holds with regard to the distances between nodes. This means that the shortest possible distance between any two nodes $u,v \in V$ in any subgraph of $H$ is always the distance of their direct connection, i.e., the weight of their edge $w(u,v)$. We use this property heavily throughout this proof. To begin with, this implies that the distance cost of $OPT$ is at least $n \cdot \sum_{u \in V}w(v^\ast,u)$. From now on, we denote $\sum_{u \in V}w(v^\ast,u)$ as $D_{v^\ast}$ to use it as a baseline distance unit. By \Cref{min_dist_bound}, it holds for any agent~$u \in V$ that the distance cost of $G$ is at most $2(n-1) d_G(u,V)$. Thus, it suffices to show that there exists some agent~$u \in V$ with $d_G(u,V) \in \mathcal{O}(\sqrt{\alpha}D_{v^\ast})$ to prove the desired distance cost ratio of $$\mathcal{O}\left( \frac{2(n-1)\sqrt{\alpha}D_{v^\ast}}{nD_{v^\ast}} \right) = \mathcal{O} \left( \sqrt{\alpha} \right).$$

    Assume towards a contradiction that $G$ is in BSE and for all nodes $u \in V$, we have $$d_G(u,V) \geq 190\sqrt{\alpha}D_{v^\ast}.$$ We proceed to show that there exists an improving move for some coalition $\Gamma \subseteq V$. To do this, we first partition $V$ into three sets of agents. Let $N$ be the set of all agents $u \in V$ with $w(u,v^\ast) \leq \frac{2}{n}D_{v^\ast}$. By the definition of $D_{v^\ast}$, we know that there exist at least $\frac{n}{2}$ agents $u \in V$, with $w(u,v^\ast) \leq \frac{2}{n}D_{v^\ast}$, because otherwise the edge weights of agents, where $w(u,v^\ast) > \frac{2}{n}D_{v^\ast}$ holds, would sum up to more than $D_{v^\ast}$. The second set of the partition is denoted as $R$ and contains all agents $u \in V$ with $w(u, v^\ast) > \frac{2}{\sqrt{\alpha}}D_{v^\ast}$. By the same argument, this set contains at most $\frac{\sqrt{\alpha}}{2}$ many agents. The set $M \coloneqq V \setminus (N \cup R)$ includes all the remaining agents of $G$. Note that $M$ as well as $R$ can be empty sets. However, if one of the sets is empty, we just assume for all $u \in V$ that $d_G(u, M) = 0$ or $d_G(u, R) = 0$, respectively. Furthermore, if $M$ is an empty set, it suffices to bound the distance cost in $N$ and $R$. Any consideration towards set $M$ and the construction in \Cref{fig:M-BGNCG_dist_ratio_BSE_proof_sketch} can then be disregarded.

    Next, we consider the distances in $G$ within $N$. Assume towards a contradiction that there exists no node $u \in N$ with $d_G(u,N) \leq 13\sqrt{\alpha}D_{v^\ast}$. Let $G_T$ be the resulting graph after all nodes in $N$ buy edges to build an almost complete $\left( \left\lfloor \frac{3n}{\sqrt{\alpha}} \right\rfloor -1 \right)$-ary tree rooted at some node $u \in N$, with a depth of at most $\log_{\left\lfloor \frac{3n}{\sqrt{\alpha}} \right\rfloor -1}(n)$. Since we have metric edge weights, we know for all $u,v \in N$ that
    $$w(u,v) \leq w(u,v^\ast) + w(v^\ast,v) \leq \frac{4}{n}D_{v^\ast}.$$
    Thus, the distance between two nodes $u,v \in N$ in $G_T$ is
    $$d_{G_T}(u,v) \leq 2\log_{\left\lfloor \frac{3n}{\sqrt{\alpha}} \right\rfloor -1}(n) \cdot \frac{4}{n}D_{v^\ast}.$$
    As any node in $N$ buys at most $\left\lfloor \frac{3n}{\sqrt{\alpha}} \right\rfloor $ edges, the cost difference between $G$ and $G_T$ for an arbitrary node $u \in N$ can be bounded as follows:
    \begin{align}
        cost(u,G) - cost(u,G_T) &\geq d_G(u,N) - \left( d_{G_T}(u,N) + \left\lfloor \frac{3n}{\sqrt{\alpha}} \right\rfloor \cdot \alpha \frac{4}{n}D_{v^\ast} \right) \\
        &\geq 13\sqrt{\alpha}D_{v^\ast} - \left( d_{G_T}(u,N) + 12\sqrt{\alpha}D_{v^\ast}\right) \\
        &\geq \sqrt{\alpha}D_{v^\ast} - d_{G_T}(u,N) \\
        \label{dist_in_N_log}
        &= \sqrt{\alpha}D_{v^\ast} - 8\log_{\left\lfloor \frac{3n}{\sqrt{\alpha}} \right\rfloor -1}(n)D_{v^\ast}.
    \end{align}
    We now show that for sufficiently large $n$ the difference in \Cref{dist_in_N_log} is always positive if $\alpha \in \omega(1)$ and $\alpha \leq n^2$. For $\alpha \leq n$, the logarithm in \Cref{dist_in_N_log} is at most
    $$8\log_{\left\lfloor 3\sqrt{n} \right\rfloor -1}(n) \leq 8\log_{\sqrt{n}}(n) = 16.$$ For $n \leq \alpha \leq n^2$ it can be upper bounded by $8\log_2(\alpha)$. Both, $16$ and $8\log_2(\alpha)$, are less than $\sqrt{\alpha}$ for $\alpha \in \omega(1)$ and sufficiently large $n$. Hence, the difference in \Cref{dist_in_N_log} is positive for all relevant ranges of $\alpha$ and building $G_T$ induces a cost reduction for all nodes in $N$. Consequently, we found a coalition $\Gamma = N$ with an improving move, which contradicts our assumption. Thus, there exists at least one agent $z \in N$ in $G$ with $d_G(z,N) \leq 13\sqrt{\alpha}D_{v^\ast}$. This also implies that there exist at least $\frac{|N|}{2} \geq \frac{n}{4}$ agents $u_i \in N$ with $d_G(u_i,z) \leq \frac{52\sqrt{\alpha}}{n}D_{v^\ast}$, since the sum of the distances to $z$, from agents $u \in N$, where $d_G(u,z) > \frac{52\sqrt{\alpha}}{n}D_{v^\ast}$ holds, would otherwise exceed $13\sqrt{\alpha}D_{v^\ast}$. We denote the set of these $\frac{n}{4}$ or more agents as $N'$.
    
    Consider set $M$ of the partition. We know for all agents $v \in M$ that
    $$\frac{2}{n}D_{v^\ast} < w(v,v^\ast) \leq \frac{2}{\sqrt{\alpha}}D_{v^\ast}.$$ We now analyze when it is beneficial for an agent~$v_i \in M$ to buy an edge towards some agent $u_j \in N'$. We call the modified network with the new edge $G_e$. Assume $d_G(v_i,z) \geq 88\sqrt{\alpha} \cdot w(v_i,v^\ast)$. We proceed to show that $v_i$ has a lower total cost in $G_e$ than in $G$. Using the triangle inequality, we lower bound the distance between agent $v_i$ and some agent $u_k \in N'$ in $G$ as follows:
    \begin{align*}
        d_G(v_i, u_k) \geq d_G(v_i, z) - d_G(z, u_k) 
        \geq 88\sqrt{\alpha} \cdot w(v_i,v^\ast) - \frac{52\sqrt{\alpha}}{n}D_{v^\ast}.
    \end{align*}
    Again using the triangle inequality we bound their distance in $G_e$:
    \begin{align*}
        d_{G_e}(v_i, u_k) &\leq d_{G_e}(v_i, z) + d_{G_e}(z, u_k) \\
        &\leq d_{G_e}(v_i, u_j) + d_G(u_j, z) + d_G(z, u_k) \\
        &\leq w(v_i,u_j) + d_G(u_j, z) + d_G(z, u_k) \\
        &\leq w(v_i,v^\ast) + w(v^\ast,u_j) + d_G(u_j, z) + d_G(z, u_k) \\
        &\leq w(v_i,v^\ast) + \frac{2}{n}D_{v^\ast} + \frac{52\sqrt{\alpha}}{n}D_{v^\ast} + \frac{52\sqrt{\alpha}}{n}D_{v^\ast} \\
        &= w(v_i,v^\ast) + \frac{104\sqrt{\alpha}+2}{n}D_{v^\ast}.
    \end{align*}
    This results in the following distance reduction between agents $v_i$ and $u_k$:
    \begin{align*}
        d_G(v_i, u_k) - d_{G_e}(v_i, u_k) &\geq 88\sqrt{\alpha} \cdot w(v_i,v^\ast) - \frac{52\sqrt{\alpha}}{n}D_{v^\ast} - \left( w(v_i,v^\ast) + \frac{104\sqrt{\alpha}+2}{n}D_{v^\ast} \right) \\
        &= (88\sqrt{\alpha}-1) w(v_i,v^\ast) - \frac{156\sqrt{\alpha}+2}{n}D_{v^\ast} \\
        &\geq (88\sqrt{\alpha}-1)\frac{2}{n}D_{v^\ast} - \frac{156\sqrt{\alpha}+2}{n}D_{v^\ast} \\
        &= \frac{20\sqrt{\alpha}-4}{n}D_{v^\ast}.
    \end{align*}
    As $u_k$ is an arbitrarily chosen node in $N'$, this distance cost reduction of agent $v_i$ after building an edge to agent $u_j$ holds towards all agents in $N'$. Thus, the total cost benefit of agent $v_i$ after buying that edge is at least
    \begin{align*}
        cost(v_i, G) - cost(v_i, G_e) &\geq \frac{n}{4} \cdot \left( d_G(v_i, u_k) - d_{G_e}(v_i, u_k) \right) - \alpha \cdot w(v_i, u_j) \\
        &\geq \frac{n}{4} \cdot \frac{20\sqrt{\alpha}-4}{n}D_{v^\ast} - \alpha \cdot (w(v_i,v^\ast) + w(v^\ast, u_j)) \\
        &\geq (5\sqrt{\alpha} - 1) D_{v^\ast} - \alpha \cdot 2w(v_i,v^\ast) \\
        &\geq (5\sqrt{\alpha} - 1) D_{v^\ast} - \alpha \cdot \frac{4}{\sqrt{\alpha}} D_{v^\ast} \\
        &= (\sqrt{\alpha} - 1) D_{v^\ast}.
    \end{align*}
    Since $\alpha \in \omega(1)$, this means building an edge to an arbitrary agent $u_j \in N'$ is beneficial for agent $v_i$. However, agent $u_j$ does not necessarily want to buy that edge.
    
    This brings us back to the beginning of the proof. We assumed that for all agents $u \in V$, we have $d_G(u,V) \geq 190\sqrt{\alpha}D_{v^\ast}$. Finally, we can define the coalition $\Gamma \subseteq V$ and its improving move in $G$. Every agent $v_i \in M$ with $d_G(v_i,z) \geq 88\sqrt{\alpha} \cdot w(v_i,v^\ast)$ buys exactly one edge to an agent in $N'$, so that every agent in $N'$ has to buy at most two edges. This is possible since $|N'| \geq \frac{n}{4} \geq \frac{|M|}{2}$. We denote the subset of agents in $M$ that buy edges as $M' \subseteq M$ and the resulting network as $G_Z$. We already established that all agents $v_i \in M'$ benefit from this change, because they get at least the same distance cost reduction as in $G_e$ from building their own edge. This only leaves the cost difference of agents in $N'$ to evaluate. The distance cost of some agent $u_j \in N'$ in $G_Z$ is at most
    \begin{align*}
        d_{G_Z}(u_j, V) &\leq n \cdot d_{G_Z}(u_j, z) + d_{G_Z}(z, V) \\
        &\leq n \cdot \frac{52\sqrt{\alpha}}{n}D_{v^\ast} + d_{G_Z}(z, V) 
        = 52\sqrt{\alpha}D_{v^\ast} + d_{G_Z}(z, V).
    \end{align*}
    We split the calculation of the distance cost of agent~$z$ in ${G_Z}$ according to the partition of $V$. The distance of agent~$z$ to nodes in $N$ is at most $13\sqrt{\alpha}D_{v^\ast}$ as we have previously shown. The distance cost to nodes in $R$ can be bounded by bounding the distance between $v^\ast$ and any other node $u \in V$. Let $P_{uv^\ast} = (u_0, u_1, \dotsc, u_k)$ be the shortest path between $u$ and $v^\ast$, where $u_0=u$ and $u_k=v^\ast$. Analogously to the proof of \Cref{lem:path_length}, the length of that path can be bounded using the triangle inequality as follows:
    \begin{align*}
        d_{G_Z}(u_0,u_k) &\leq \sum_{i=0}^{k-1} w(u_i, u_{i+1}) \\
        &\leq \sum_{i=0}^{k-1} \left( w(u_i, v^\ast) + w(u_{i+1}, v^\ast) \right) 
        \leq 2D_{v^\ast}.
    \end{align*}
    Since $d_{G_Z}(z,u) \leq d_{G_Z}(z,v^\ast) + d_{G_Z}(v^\ast,u)$, this implies that $d_{G_Z}(z,u) \leq 4D_{v^\ast}$. Furthermore, the set $R$ contains at most $\frac{\sqrt{\alpha}}{2}$ nodes. Thus, it holds that
    $$d_{G_Z}(z,R) \leq \frac{\sqrt{\alpha}}{2} \cdot 4D_{v^\ast} = 2\sqrt{\alpha}D_{v^\ast}.$$
    By definition of $M'$, the distance of agent $z$ to nodes in $M \setminus M'$ is
    \begin{align*}
        d_{G_Z}(z,M \setminus M') &\leq \sum_{v \in M \setminus M'} 88\sqrt{\alpha} \cdot w(v,v^\ast) \\
        &= 88\sqrt{\alpha} \sum_{v \in M \setminus M'} w(v,v^\ast)\leq 88\sqrt{\alpha}D_{v^\ast}.
    \end{align*}
    Lastly, the distance to nodes in $M'$ can be bounded by considering the new edge each node in $M'$ buys. Let $B$ be the set of newly built edges in $G_Z$. Assume for each of the following edges $\{u_i,v_j\} \in B$ that $u_i \in N'$. As each agent in $N'$ or $M'$ owns at most two edges in $B$, we can use the triangle inequality and the definition of $D_{v^\ast}$ and have
    \begin{align*}
        d_{G_Z}(z,M') &\leq \sum_{\{u_i,v_j\} \in B} d_{G_Z}(z,u_i) + w(u_i, v_j) \\
        &\leq \sum_{\{u_i,v_j\} \in B} \frac{52\sqrt{\alpha}}{n}D_{v^\ast} + w(u_i,v_j) \\
        &\leq |B| \cdot \frac{52\sqrt{\alpha}}{n}D_{v^\ast} + \sum_{\{u_i,v_j\} \in B} w(u_i,v_j) \\
        &\leq \frac{n}{2} \cdot \frac{52\sqrt{\alpha}}{n}D_{v^\ast} + \sum_{\{u_i,v_j\} \in B} \left( w(u_i,v^\ast) + w(v^\ast, v_j) \right) \\
        &\leq 26\sqrt{\alpha}D_{v^\ast} + 2D_{v^\ast} \\
        &= (26\sqrt{\alpha}+2)D_{v^\ast}.
    \end{align*}
    Summing everything up, this results in a total distance cost of agent $z$ in $G_Z$ of
    \begin{align*}
        d_{G_Z}(z, V) &= d_{G_Z}(z, N) + d_{G_Z}(z, M \setminus M') + d_{G_Z}(z, M') + d_{G_Z}(z, R) \\
        &\leq 13\sqrt{\alpha}D_{v^\ast} + 88\sqrt{\alpha}D_{v^\ast} + (26\sqrt{\alpha}+2)D_{v^\ast} + 2\sqrt{\alpha}D_{v^\ast} \\
        &= (129\sqrt{\alpha}+2)D_{v^\ast}
    \end{align*}
    Regarding the total distance cost of some agent $u_i \in N'$ in $G_Z$, this implies that
    $$d_{G_Z}(u_i, V) \leq 52\sqrt{\alpha}D_{v^\ast} + (129\sqrt{\alpha}+2)D_{v^\ast} = (181\sqrt{\alpha}+2)D_{v^\ast}.$$
    Since each agent in $N'$ buys at most two new edges to agents in $M'$, the total cost benefit for agent $u_i$ is at least
    \begin{align*}
        cost(u_i,G) - cost(u_i,G_Z) &\geq d_G(u_i, V) - \left( d_{G_Z}(u_i, V) + 2\alpha \cdot \frac{4}{\sqrt{\alpha}}D_{v^\ast} \right) \\
        &\geq 190\sqrt{\alpha}D_{v^\ast} - \left( (181\sqrt{\alpha}+2)D_{v^\ast} + 8\sqrt{\alpha}D_{v^\ast} \right) \\
        &= (\sqrt{\alpha}-2)D_{v^\ast}.
    \end{align*}
    This means all agents in $N'$ and all agents in $M'$ profit by the suggested change to $G_Z$ if $\alpha \in \omega(1)$. Thus, we have found a coalition $\Gamma = N' \cup M'$ with an improving move and $G$ is not in BSE. This contradicts our initial assumption. Hence, there exists an agent $u \in V$ with $d_G(u,V) \leq 190\sqrt{\alpha}D_{v^\ast}$, which is in $\mathcal{O}(\sqrt{\alpha}D_{v^\ast})$. Using \Cref{min_dist_bound}, we can finally bound the distance cost ratio as follows:
    \begin{align*}
        \frac{\sum_{u \in V} d_G(u,V)}{\sum_{v \in V} d_{OPT}(v,V)} &\leq \frac{2(n-1) \cdot 190\sqrt{\alpha}D_{v^\ast}}{n \cdot D_{v^\ast}} 
        \leq 380\sqrt{\alpha}.
    \end{align*}
    Therefore, in the M-BGNCG, for $\alpha \in \omega(1)$ and $\alpha \leq n^2$ the distance cost ratio between any BSE $G = (V,E)$ and $OPT$ is in $\mathcal{O}(\sqrt{\alpha})$. 
\end{proofE}

Combining \Cref{M-BGNCG_dist_cost_ratio_BSE} with \Cref{edge_cost_bound_by_dist_cost} directly yields a bound on the price of anarchy. In particular, this bound demonstrates that for the M-BGNCG with $\alpha\in o(n^{4/3})$ there is a level of cooperation for which a lower price of anarchy compared to pairwise stable networks is guaranteed.

\begin{corollaryE}[][normal, category=metric]
    \label{PoA_upper_M-BGNCG_BSE}
    For $\alpha \leq n^2$, the price of anarchy of the M-BGNCG with respect to bilateral strong equilibria is in $\mathcal{O} \left( \max \left\{ \sqrt{\alpha}, \frac{\alpha \sqrt{\alpha}}{n} \right\} \right)$.
\end{corollaryE}

 For $\alpha \in \mathcal{O}(n)$, this results in a price of anarchy in $\mathcal{O} \left( \sqrt{\alpha} \right)$ for M-BGNCG networks in BSE, which is asymptotically tight due to the fourth lower bound in \Cref{M-BGNCG_PoA_lower_bounds}. For the ranges of $\alpha$ where our bounds are not asymptotically tight, we suspect that the actual price of anarchy for M-BGNCG networks in BSE is much closer to the lower bounds in \Cref{M-BGNCG_PoA_lower_bounds} than to the upper bound in \Cref{PoA_upper_M-BGNCG_BSE}. There are two main reasons for this. The first is that there is probably a better upper bound on the total edge cost of the network in BSE than the bound given in \Cref{edge_cost_bound_by_dist_cost}, which just bounds the total edge cost of the network dependent on its total distance cost. The second reason is that a significantly higher total distance cost of a network in BSE could imply a higher total edge cost in its corresponding social optimum network. In particular, we conjecture the following:
 \begin{ourconjecture}
     Let $G=(V,E)$ be a M-BGNCG network in BSE and let $OPT=(V,E_{OPT})$ be the corresponding social optimum network. Furthermore, let the total distance cost of $G$ be in $\Omega \left( 2(n-1)\alpha^xD_{v^\ast} \right)$, where $D_{v^\ast}$ is defined as in \Cref{M-BGNCG_dist_cost_ratio_BSE}, and $x \in \R_0^+$. Then the edge cost of $OPT$ is in $\Omega \left( \alpha^{2x}D_{v^\ast} \right)$.
 \end{ourconjecture}
 This would imply the following conjectured price of anarchy for M-BGNCG networks in BSE:
 \begin{align}
     \frac{cost(G)}{cost(OPT)} &\in \mathcal{O} \left( \frac{\alpha \cdot w(E) + \sum_{u \in V} d_G(u,V)}{\alpha \cdot w(E_{OPT}) + \sum_{u \in V} d_{OPT}(u,V)} \right)\nonumber \\
     \label{conjectured_PoA}
     &= \mathcal{O} \left( \frac{n\sqrt{\alpha}}{\alpha + n} + \frac{\alpha \cdot w(E)}{\alpha \cdot w(E_{OPT}) + \sum_{u \in V} d_{OPT}(u,V)} \right).
 \end{align}
 The left part of the sum in \Cref{conjectured_PoA} matches our fourth lower bound in \Cref{M-BGNCG_PoA_lower_bounds}. For the edge cost ratio between a M-BGNCG network in BSE and its corresponding social optimum network, we conjecture an upper bound of roughly $\mathcal{O} \left( \max \left\{ \log(\alpha), \frac{\alpha}{n} \right\} \right)$. This would result in all bounds on the price of anarchy for M-BGNCG networks in BSE being asymptotically tight. The price of anarchy for $\alpha \in \mathcal{O}(n^2)$ would then be $\Theta ( \frac{n\sqrt{\alpha}}{\alpha + n} + \frac{\alpha}{n} )$.
 
Also, we do not provide a better upper bound on the price of anarchy for M-BGNCG networks in BNE than the bounds for pairwise stable networks in \Cref{PoA_upper_M-BGNCG_PS_alpha} and \Cref{PoA_upper_M-BGNCG_PS_n}. We expect the price of anarchy for these networks to be in $\Theta \left( \sqrt{\alpha} \right)$ or at least in $o(\alpha)$.

\section{Conclusion and Outlook}
We study the impact of cooperation on the price of anarchy in the bilateral version of the generalized network creation game~\cite{GNCG}. For this, we consider the solution concepts pairwise stability, bilateral neighborhood equilibria, and bilateral strong equilibria, that allow for increasingly more cooperation among the agents. We find that even allowing the strongest form of cooperation does not yield an improved price of anarchy on arbitrarily weighted host graphs. In contrast, if the host graph has metric edge weights and if $\alpha \in o(n^{4/3})$, then strong cooperation is guaranteed to yield a significantly lower price of anarchy. Thus, both ingredients, strong cooperation and metric weights, are necessary.

In particular, with our novel proof technique, we show that for $\alpha \in \mathcal{O}(n)$, in the M-BGNCG, the lower and upper bounds on the price of anarchy of networks in BSE are asymptotically tight, resulting in a bound of $\Theta \left( \sqrt{\alpha} \right)$. Moreover, for pairwise stability, the price of anarchy is always at least linear in $\alpha$ or in $n$, even with metric weights. Therefore, a high price of anarchy might only be prevented by allowing the cooperation of $\omega(1)$ agents and we believe that actually cooperation of $\Omega(n)$ agents is needed. Note that this is in stark contrast to the results for equilibrium networks that are trees in the BNCG by Friedrich, Gawendowicz, Lenzner, and Zahn~\cite{CooperationBNCG}, where the cooperation of at most three agents already suffices to ensure a constant PoA.   
In practice, for policy-makers and governing network operators, the need for larger coalitions yields interesting design and implementation questions of how to initiate, coordinate, and to sustain such agent cooperations.

We conjecture that some of our PoA upper bounds can be improved. For the M-BGNCG, for $\alpha \in \mathcal{O}(n^2)$, we conjecture that the PoA with respect to BSE actually is in $\Theta \big( \frac{n\sqrt{\alpha}}{\alpha + n} + \frac{\alpha}{n} \big)$. Furthermore, for M-BGNCG networks in BNE, we only provide the same upper bound on the PoA as for networks in PS, which is a weaker form of cooperation. We conjecture that with the BNE there should be an asymptotically lower upper bound on the PoA, maybe even as low as $\mathcal{O} \left( \sqrt{\alpha} \right)$. We are positive that our novel techniques and structural observations might be helpful for progress on this. 

Last but not least, we emphasize that the existence of (approximate) equilibria is a challenging open problem for all network creation models with weighted edges. So far, only positive results for very restricted weighted host graphs are known, and a complete characterization is missing.

%
%
%
\bibliographystyle{splncs04}
\bibliography{references.bib}

\end{document}